\documentclass[7pt, two column]{wlscirep} 
\usepackage[utf8]{inputenc} 
\usepackage[T1]{fontenc}
\usepackage[super,square,comma,sort&compress]{natbib}
\usepackage{bm} 
\usepackage{xr} 
\usepackage[version=3]{mhchem}    
\usepackage{balance}
\usepackage{mathptmx}
\usepackage{graphicx} 
\usepackage{pgfplots}
\usepackage{subcaption}
\usepackage{lastpage}   

\usepackage{float}
\usepackage{fancyhdr}
\usepackage{array}
\usepackage{droidsans}
\usepackage{charter}
\usepackage[T1]{fontenc}
\usepackage{epstopdf}

\usepackage{epstopdf,epsfig}
\usepackage{bm}
\usepackage{authblk}
\usepackage{overpic} 
\usepackage{tabularx}  
\usepackage{booktabs}   
\usepackage{array}     

\usepackage{mathtools}  

\usepackage{enumitem}

\DeclareCaptionLabelFormat{bflabel}{\textbf{Fig. #2 | }}
\DeclareCaptionLabelFormat{bftable}{\textbf{Table #2 | }}  
\usepackage{lettrine}

\usepackage{placeins}

\pgfplotsset{                            
    every axis/.append style={           
    line width=0.25pt, 
        tick style={line width=0.25pt} 
        }}

\newlength\figH
\newlength\figW
\usepackage{adjustbox} 
\pgfplotsset{compat=newest}

\definecolor{cream}{RGB}{222,217,201}

\makeatletter

\renewcommand\normalsize{%
  \@setfontsize\normalsize{8pt}{10pt}
}
\normalsize 

\renewcommand\small{%
  \@setfontsize\small{7pt}{9pt}
}
\renewcommand\footnotesize{%
  \@setfontsize\footnotesize{6pt}{8pt}
}
\makeatother

\title{
Droplet Impact on Microparticle Raft:
Wettability, density and size govern splashing and microplastic ejection from rafts under raindrop impact.}

\author[1]{\textit{Muhammad Hamza Iqbal}}
\author[2]{\textit{Alfonso Arturo Castrej\'on-Pita}}
\author[1$^\dagger$]{\textit{Jos\'e Rafael Castrej\'on-Pita}}
\author[3*]{\textit{Miguel A. Quetzeri-Santiago}}

\affil[1]{Department of Mechanical Engineering, University College London, WC1E 7JE, London, United Kingdom}
\affil[2]{Department of Engineering Science, University of Oxford, Oxford, OX1 3PJ, Uniied Kingdom} 
\affil[3]{Instituto de Investigaciones en Materiales, Universidad Nacional Autónoma de México, Cd. Universitaria, Mexico City, 04510,Mexico}

\affil[$^\dagger$]{e-mail: r.pita@ucl.ac.uk}
\affil[*]{e-mail: mquetzeri@materiales.unam.mx }

\begin{abstract}
\normalsize{

Raindrop impact on the ocean has been proposed as a mechanism for microplastic transfer from seawater to the atmosphere, yet the interfacial dynamics governing particle ejection from floating microplastics remain largely unexplored. We investigate droplet impact onto microparticle monolayers (rafts) spanning a wide range of sizes, contrasting densities, and wettabilities, under raindrop-relevant impact conditions. Particle rafts strongly influence splash dynamics, cavity collapse, and Worthington jet formation. Splash onset is controlled by particle-induced roughness and capillary adhesion: deeply immersed particles stabilise the spreading lamella, producing only microdroplets, whereas weakly immersed particles destabilise the rim, promoting fingering and splashing. Following impact, raft characteristics govern wave–swell dynamics, separating into elastic and rigid regimes. Superhydrophobic rafts enable particle ejection upon impact and form particle-armoured Worthington jets that fragment into liquid marbles, providing an efficient aerosolisation pathway. In contrast, less hydrophobic rafts show limited detachment upon impact but still support Worthington jet-mediated transport. Despite these differences, splash thresholds and Worthington jet heights collapse under a simple geometric–inertial–capillary scaling. These results show how particulate monolayers modify canonical droplet-impact and identify the interfacial conditions under which rainfall transfers microplastics from ocean to atmosphere, and inform related droplet–granular processes such as soil erosion, and impacts on sandy substrates.

}
\end{abstract}

\begin{document}
\maketitle

\section*{Introduction}

\lettrine[lines=3]{\textcolor{orange}{M}}{}icroplastics are now recognised as pervasive contaminants in both the hydrosphere and atmosphere alike, with substantial land-to-sea inputs and vast surface inventories documented across the global ocean \cite{jambeck2015plastic, strokal2023river, eriksen2014plastic, cozar2014plastic, pabortsava2020high}. Consistent definitions and size classifications (from sub-micrometre to <5 mm; fragments, films and fibres) have been formalised to enable cross-study comparability and reliable accounting \cite{ hartmann2019we, frias2019microplastics}. Beyond aquatic accumulation, airborne transport distributes particles to remote regions and even the free troposphere, demonstrating efficient exchange pathways between surface waters and the atmosphere \cite{allen2019atmospheric, evangeliou2020atmospheric, allen2021evidence}. 

Microplastics have also been detected in human lung tissue and bronchoalveolar lavage fluid, underscoring plausible inhalation exposure and the need to constrain emission mechanisms at the air–water boundary \cite{amato2021presence, jenner2022detection,  qiu2023evidence}. A growing body of field evidence suggests that the ocean is not merely a sink but also a source of airborne microplastics \cite{allen2020examination}. For ocean-to-atmosphere transport to occur, plastics must first accumulate at the air–sea boundary. The sea surface microlayer (SML), a $1-1000$ $\mathrm{\mu m}$ thick, biogeochemically enriched interface, has been shown to contain microplastic concentrations orders of magnitude higher than bulk seawater \cite{song2014large, anderson2018rapid}. This concentrated reservoir forms a highly susceptible pool for emissions. The dominant, well-studied atmospheric release pathway is sea aerosol generated by breaking waves, whereby bubbles scavenge particles while rising to then expel them in surface jets upon bursting \cite{masry2021experimental, shaw2023ocean}. The microplastics are transferred via two distinct mechanisms: (i) rupture of the thin bubble cap film, which produces numerous submicron film droplets capable of carrying the smallest microplastics (typically $< 1-10$ $\mathrm{\mu m}$) that become trapped within the interfacial film, and (ii) collapse of the underlying cavity, which drives a high-speed Worthington jet that breaks up into larger jet droplets capable of ejecting substantially bigger microplastics, from tens to several hundreds of micrometres, that reside at or near the bubble cap surface \cite{shaw2023ocean}. Laboratory measurements and global modelling suggest that this bubble-mediated emission pathway may contribute on the order of 0.1 million tonnes per year of microplastics to the atmosphere \cite{shaw2023ocean, oehlschlagel2024water}. Yet, this mechanism strongly depends on wind and wave conditions and does not fully resolve the pervasive presence of microplastics in atmospheric samples collected over calm seas and coastal regions.

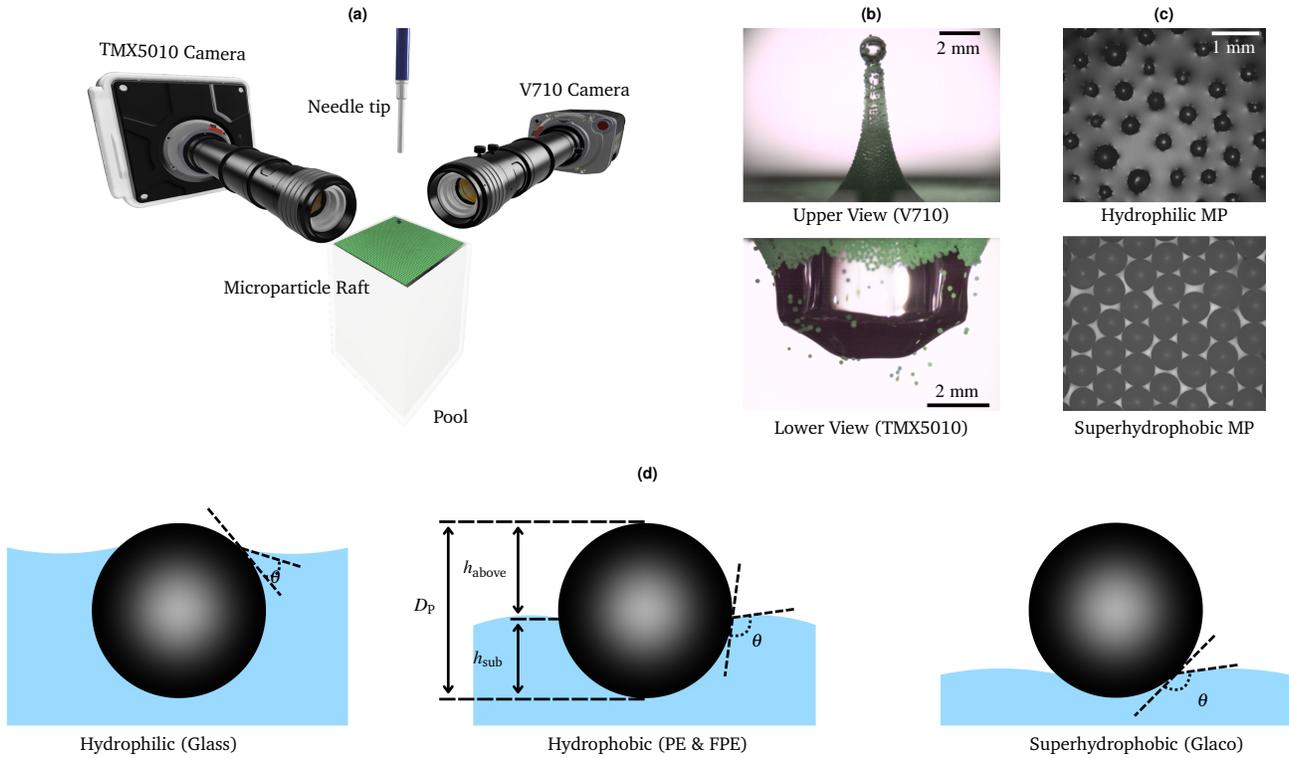
\begin{figure*}[t!]\
    \centering
    \begin{subfigure}[t]{0.54\textwidth} 
        \centering
        \vspace{-7mm}
        \caption{} 
        \begin{tikzpicture}[scale=1.0]
            \node[inner sep=0pt] (img) at (0,0) {\includegraphics[width=0.8\textwidth]{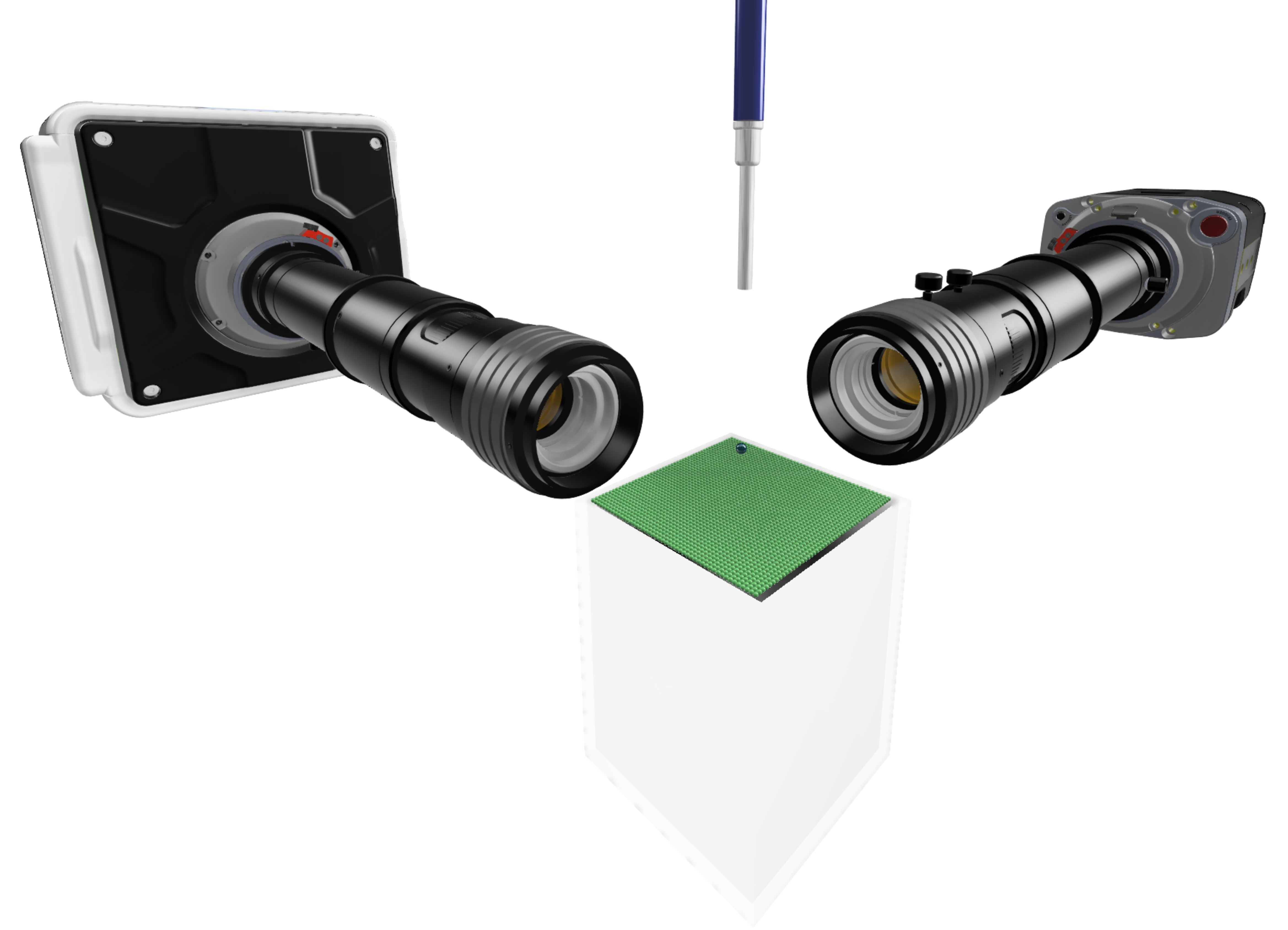}};
            
            \node[text=black, font=\scriptsize, align=left, rotate= -2] at (-2.45, 2.35) {\text{TMX5010 Camera}};
            \node[text=black, font=\scriptsize, align=left, rotate=  0] at ( -0.1,  1.60) {\text{Needle tip}}; 
            \node[text=black, font=\scriptsize, align=left, rotate=  0] at (  2.9,  1.85) {\text{V710 Camera}}; 
            \node[text=black, font=\scriptsize, align=left, rotate=  0] at (-0.80, -0.80) {\text{Microparticle Raft}}; 
            \node[text=black, font=\scriptsize, align=left, rotate=  0] at ( 1.25, -2.50) {\text{Pool}}; 
            
        \end{tikzpicture}
    \end{subfigure}
    \begin{subfigure}[t]{0.25\textwidth} 
        \centering
        \hspace{-50mm}
        \vspace{-7mm}
        \caption{} 
        \begin{tikzpicture}[scale=0.9]
            \node[inner sep=0pt] (img) at (0,0) {\includegraphics[width=0.8\textwidth]{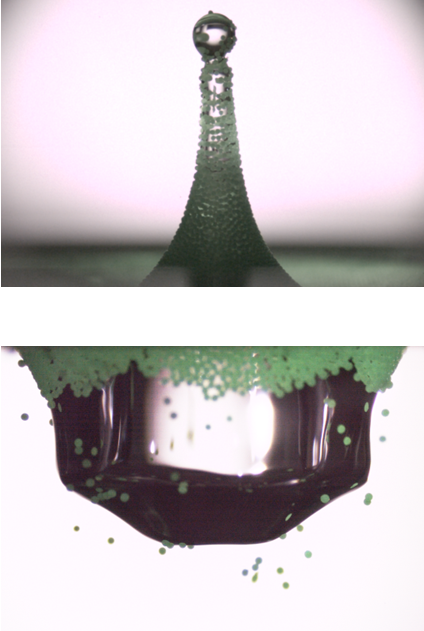}};
            
            \draw[line width=0.5mm, draw=black, -] (1.00, 2.75) -- (1.60, 2.75);
            \node[text=black, font=\scriptsize] at (1.28, 2.75-0.21) {\textbf{$2~\mathrm{mm}$}}; 
            \node[text=black, font=\small, anchor=center] at (0.00, 0.05) {\text{Upper View (V710)}};

            \draw[line width=0.5mm, draw=black, -] (0.82, -2.75) -- (1.74, -2.75);
            \node[text=black, font=\scriptsize, anchor=center] at (1.26, -2.75+0.21) {\textbf{$2~\mathrm{mm}$}}; 
            \node[text=black, font=\small, anchor=center] at (0.00, -3.10) {\text{Lower View (TMX5010)}};

        \end{tikzpicture}
    \end{subfigure}
    \begin{subfigure}[t]{0.1965\textwidth} 
        \centering
        \vspace{-7mm}
        \caption{} 
        \begin{tikzpicture}[scale=0.9]
            \node[inner sep=0pt] (img) at (0,0) {\includegraphics[width=0.8\textwidth]{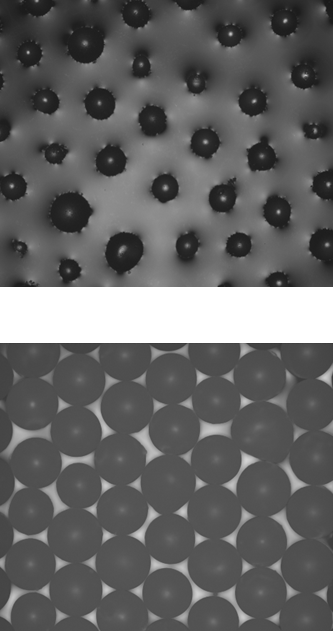}};
            
            \draw[line width=0.5mm, draw=white, -] (0.70, 2.75) -- (0.70+0.70, 2.75); 
            \node[text=white, font=\scriptsize] at (1.02, 2.55) {\textbf{$1~\mathrm{mm}$}}; 

            \node[text=black, font=\small, anchor=center] at (0.00,  0.05) {\text{Hydrophilic MP}};
            \node[text=black, font=\small, anchor=center] at (0.00, -3.10) {\text{Superhydrophobic MP}};
            
        \end{tikzpicture}
    \end{subfigure}

    \begin{subfigure}[t]{1.00\textwidth}   
        \centering
        \vspace{0mm}
        \caption{} 
        \begin{tikzpicture}[scale=1.00]
            \node[inner sep=0pt] (img) at (0,0) {\includegraphics[width=1\textwidth]{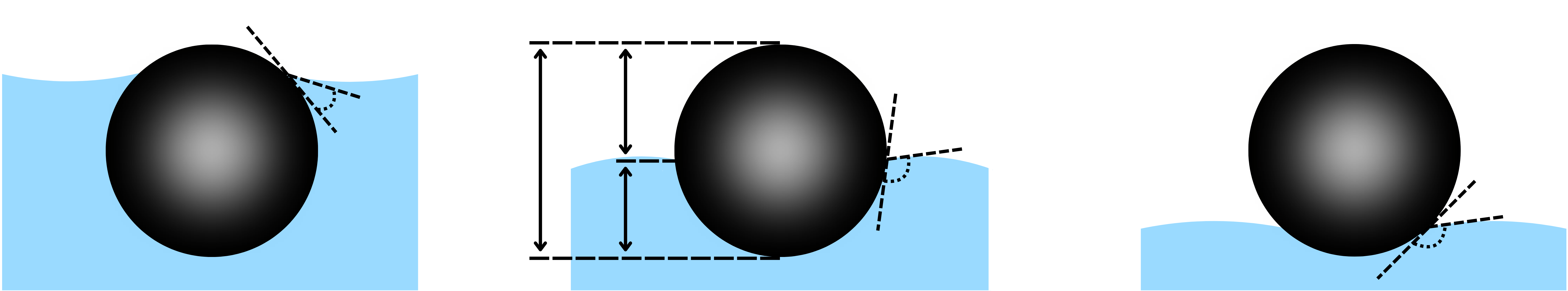}};
            \node[text=black, font=\small, anchor=center] at (-6.5, -1.85) {\text{Hydrophilic (Glass)}}; 
            \node[text=black, font=\small, anchor=center] at ( 0.0, -1.85) {\text{Hydrophobic (PE \& FPE)}}; 
            \node[text=black, font=\small, anchor=center] at ( 6.5, -1.85) {\text{Superhydrophobic (Glaco)}}; 

            
            \node[text=black, font=\scriptsize] at     (-2.16, 0.50) {$h_{\text{above}}$}; 

            \node[text=black, font=\scriptsize] at     (-2.12, -0.70) {{$h_{\text{sub}}$}}; 

            \node[text=black, font=\scriptsize] at    (-2.96, 0.00) {{$D_{\text{P}}$}}; 
            
            \node[text=black, font=\small] at         (-4.95,  0.40) {\textbf{$\theta$}}; 
            \node[text=black, font=\small] at         ( 1.45, -0.45) {\textbf{$\theta$}};
            \node[text=black, font=\small] at         ( 7.38, -1.25) {\textbf{$\theta$}}; 
        \end{tikzpicture}
    \end{subfigure}
    
    \caption{\textbf{Experimental configuration and characterisation of floating microparticle rafts.}
    \textbf{a} Sketch of the droplet–raft impact apparatus. Millimetric water droplets are generated from a syringe pump and needle and released onto a quiescent pool whose surface is coated with a closely packed particle raft. A high-speed side-view imaging system records the droplet impact, crater formation and Worthington jet dynamics.
    \textbf{b} Representative top-view (top) and side-view (bottom) images from the high-speed cameras showing the hexagonally packed monolayer of $780~\mathrm{\mu m}$ FPE microparticles and the corresponding interfacial profile at the particle scale.
    \textbf{c} Optical profilometer scans for hydrophilic $625~\mathrm{\mu m}$ glass spheres (top) and Glaco-treated $550~\mathrm{\mu m}$ FPE spheres (bottom), used to determine the emerged height ($h_\text{above}$) of the particle above the undisturbed interface.
    \textbf{d} Schematic immersion geometries for hydrophilic, hydrophobic and superhydrophobic microparticle spheres. The submerged depth ($h_\text{sub}$), emerged height ($h_\text{above}$), and the apparent contact angle ($\theta$) are indicated (see Table~\ref{table1}), corresponding to the geometric relation used to compute wettability at negligible Bond number. Not to scale. 
    }
    
    \label{figure1setup}  
\end{figure*} 

A complementary and not well-studied emission arises from the impact of raindrops on the ocean surface. Millimetric raindrops strike the ocean surface with kinetic energies on the order of $10^{-5}-10^{-4}~\mathrm{J}$, driving the expansion and lift-off of an ultra-thin ($\sim 1-10~\mathrm{\mu m}$) ejecta sheet at the neck, forming crowns that generate secondary droplets \cite{Yarin2006Drop, thoroddsen2002ejecta}. The onset and intensity of this splashing transition depend primarily on the impact conditions and the ensuing liquid–liquid coalescence and lamella dynamics, which determine whether the expanding sheet lifts off and fragments into airborne droplets \cite{Agbaglah2014growth}. Similarly, recent simulation–experiment studies have demonstrated that secondary droplets produced by rainfall impact can directly entrain small-sized ($6~\mathrm{\mu m}$) particles from the ocean surface. In fact, the particle concentration in the ejecta remains comparable to that in the surface water layer \cite{Lehmann2021ejection}. Upscaling these single-impact dynamics with global rainfall statistics suggests that this mechanism may release up to $\sim 10^{14}$ microplastic particles into the atmosphere annually \cite{Lehmann2021ejection}. Earlier aerosol studies further have shown that prompt splash on aqueous solution generates a population of micron-sized secondary droplets that can eject solid residues (up to $30~\mathrm{\mu m}$) into the air, with the quantity of this airborne material increasing with impact inertia \cite{motzkus2008study, motzkus2011study}. 

A central feature of droplet impact on a liquid pool is the emergence of a Worthington jet formed when the transient cavity created by the impacting drop collapses and focuses momentum upward along the symmetry axis \cite{Worthington1908}. This vertical jet can accelerate fluid parcels to several metres per second \cite{ray2015regimes}, projecting droplets to heights far exceeding those generated by crown splash alone. Thus, representing an efficient upward transport pathway for any material residing at or near the interface, including microplastics concentrated within the sea surface microlayer, into the air/atmosphere. 
The height, morphology, and breakup of the jet are governed by a complex interplay of inertial, capillary and gravitational forces, which define distinct jetting regimes. These force balances are captured by the Weber number, $We = \rho U^{2}D/\sigma$, which compares inertial to capillary stresses, and the Froude number, $Fr = U^{2}/gD$, which compares inertial to gravitational forces, where $\rho$, $U$, $D$, $\sigma$ and $g$ denote liquid density, impact velocity, droplet diameter, surface tension and gravitational acceleration, respectively. 
Numerical and experimental studies of drop–pool impact dynamics have shown that, as the Weber and Froude numbers increase, the collapsing cavity transitions systematically through U-shaped, V-shaped, and W-shaped profiles, each producing a distinct jet type, from short thick jets to high-speed slender jets and long thick jets, respectively, with correspondingly different pinch-off behaviours \cite{ray2015regimes, michon2017jet}. These transitions arise from two underlying jetting pathways, namely a thin, rapidly rising "singular jet" driven by the focused collapse of a capillary wave at intermediate We/Fr and a slower, thicker "cavity jet" produced by the large-scale retraction of the crater at higher We/Fr \cite{michon2017jet}. 
Furthermore, increasing impact velocity leads to a well-defined regime progression, where the ejected secondary droplet can not only be comparable to, but even exceed, the size of the original impacting drop, underscoring the capacity of this mechanism to expel substantial volumes of surface water \cite{das2022evolution}. 
Whether the jet fragments into airborne droplets is governed by the critical balance between viscous and capillary–inertial stresses, captured by the Ohnesorge number $Oh = \mu/\sqrt{\rho \sigma D}$, where $\mu$ denote liquid viscosity. Above a threshold Ohnesorge number ($Oh \approx 0.09$), viscous damping suppresses Rayleigh–Plateau jet breakup entirely, even at high impact velocities \cite{Castillo2015droplet}.
These dynamics are sensitive to surface topology; impacts on pre-existing wave troughs, for instance, can triple the jet height and increase the total ejected liquid volume by over sixfold compared to a flat surface, enhancing aerosolisation \cite{Singh2022droplet}. 
In addition, the splashing and jet dynamics are altered when the free surface is no longer a bare (clean) liquid but is covered by a closely packed monolayer: a “particle raft.” Particles residing at the liquid–air interface deform the meniscus according to a balance between gravity and surface tension, captured by the Bond number $Bo = We/Fr=\rho g D^{2}/\sigma$ \cite{protiere2023particle}. For particles with $Bo \ll 1$, surface tension dominates over weight, and even small interfacial slopes can generate long-ranged capillary interactions that drive particles to migrate and cluster. These interactions, classically described in studies of capillary attraction between floating bodies \cite{nicolson1949interaction, stamou2000long}, and in the broader framework of interfacial colloidal assembly \cite{kralchevsky2000capillary}, cause particles with similar meniscus curvature to attract while those with opposite curvature repel, leading to the Cheerios effect \cite{vella2005cheerios}. Once trapped at the interface, particles experience attachment energy arising from the replacement of a patch of fluid–fluid interface by particle–fluid interfaces and form mechanically stable monolayers whose structural and elastic properties have been widely characterised \cite{Binks2006colloidal, kralchevsky2001particles}. When monodisperse particles with uniform wettability accumulate at an interface, capillary forces draw them into contact and favour dense hexagonal packing that minimises interfacial energy, producing a mechanically coherent, quasi-solid raft. \cite{Binks2006Section1, Binks2006Section2}. Such monolayers exhibit a dual mechanical character: on macroscopic scales they respond as an elastic sheet, yet on the particle scale they behave as a granular assembly capable of rearrangement, buckling and fracture under stress, depending on surface coverage and confinement \cite{protiere2023particle}. 

Here, we investigate droplet impact dynamics onto quiescent pools covered with monodisperse particle rafts. In each experiment, a single particle type (defined by one size, density, and wettability) is used to form a uniform monolayer raft, enabling systematic variation of interfacial properties across fluorescent polyethylene, polyethylene (FPE and PE) and glass in both untreated and Glaco-treated superhydrophobic states. Orthogonal high-speed imaging resolves the spatiotemporal evolution of the ejecta sheet, cavity and the Worthington jet, allowing us to quantify jet height, breakup pathways and the conditions governing microplastic ejection from the interface.
Together, these measurements provide the first direct characterisation of the role of particle rafts on the dynamics of drop–pool impact and, consequently, the mechanisms governing atmospheric release of microplastics from the ocean surface.

\begin{table}[t]
\centering
\noindent\rule{\columnwidth}{0.1pt}
\caption{\textbf{Physical properties of microparticles. }}
\small
\setlength{\tabcolsep}{4pt} 

\resizebox{\columnwidth}{!}{%
\begin{tabular}{lccc}
\hline

$\textbf{Microparticle}$ & ${\rho_p~\mathrm{(g\,cm^{-3})}}$ & $h$ & ${\theta}~ \mathrm{(^\circ)}$ \\

\hline 

FPE (untreated)   & 1.02 & 0.41 $\pm$ 0.03 &     100 $\pm$ 5 \\
G-FPE (treated)   & 1.02 & 0.04 $\pm$ 0.02 &     157 $\pm$ 3 \\
PE (untreated)    & 1.35 & 0.43 $\pm$ 0.02 & \ \ 98 $\pm$ 4 \\
G-PE (treated)    & 1.35 & 0.06 $\pm$ 0.03 &     152 $\pm$ 4 \\
Glass (untreated) & 2.50 & 0.90 $\pm$ 0.03 & \ \ 37  $\pm$ 4 \\
G-Glass (treated) & 2.50 & 0.11 $\pm$ 0.04 &     141 $\pm$ 6 \\

\hline
\end{tabular}}

\vspace{2pt}
\noindent\parbox{\columnwidth}{
\raggedright
\footnotesize
$\rho_p$ is the particle density, $h$ is the dimensionless immersion ratio ($h_{sub}/D_p$), and $\theta$ denotes the contact angle calculated from $h$ (see Methods). Values are reported as mean $\pm$ standard deviation.}

\label{table1}

\end{table}

\section*{Results} 

In our experiments, we studied the effect of particle size and wettability on the resulting dynamics following droplet impact on rafts. 
Figure~\ref{figure1setup} illustrates the experimental setup (a), shows an example of the impact dynamics (b), the typical raft geometries prior impact (c), and defines the experimental parameters (d). In brief, a $2.55\,\mathrm{mm}$ water droplet is generated from a syringe pump and released onto a quiescent pool whose surface is covered with a close-packed monolayer of monodisperse microparticles (Fig.~\ref{figure1setup}a). Orthogonal high-speed imaging captures the resulting crater/cavity formation, the cavity collapse and the resulting Worthington-jet dynamics (Fig.~\ref{figure1setup}b).
The droplet–raft impact experiments are dependent on the interfacial mechanics of the microparticles within the floating monolayer raft. Prior the impact, the microparticles self-assemble into uniform hexagonal rafts with their immersion geometries characterised using vertical-scanning profilometry (Fig.~\ref{figure1setup}c and Methods). These measurements provide the emerged height above the pool level $h_{above}$ and the submerged depth $h_{sub}$ for each particle type. We note that a characteristic contact angle is well defined at the gas-particle-pool interface, which depends on the particle wettability.  We compute the dimensionless immersion ratio as $h=h_{sub}/D_{p}$, and the corresponding geometric contact angle as $\theta$ (Fig.~\ref{figure1setup}d). 
Since all the particles lie in the low Bond number regime ($Bo\ll1$) gravity does not significantly distort the interface, the immersion geometry is set by capillarity alone and $\theta$ can be obtained from geometric relationships. The resulting values of $h$ and $\theta$ vary markedly with particle material and surface treatment (Table~\ref{table1}). 

\begin{figure*}[t!]
    \begin{subfigure}[t]{0.49\textwidth}   
        \centering
        \vspace{-7mm}
        \caption{}         
        \input{Figures/Fig2_Splash_Untreated}
    \end{subfigure}
    \begin{subfigure}[t]{0.49\textwidth}   
        \centering
        \vspace{-7mm}
        \caption{}         
        \input{Figures/Fig2_Splash_treated}
    \end{subfigure}

    \begin{subfigure}[t]{1.00\textwidth}   
        \centering
        \vspace{-1mm}
        \caption{} 
        \begin{tikzpicture}[scale=1.00]
            \node[inner sep=0pt] (img) at (0,0) {\includegraphics[width=1\textwidth]{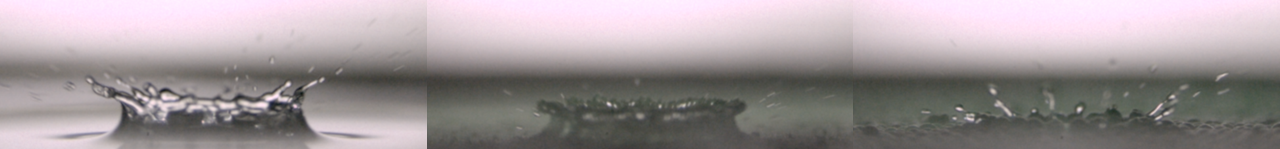}};
            
            \def\xval{7.70} 
            \def\yval{0.90}
            \draw[line width=0.5mm, draw=black, -] (\xval, \yval) -- (\xval+0.94, \yval); 
            \node[text=black, font=\footnotesize] at (\xval+0.47, \yval-0.19) {\textbf{2 mm}}; 
            
            \node[text=black, font=\small, anchor=center] at (-6.0, -1.25) {\text{Water}}; 
            \node[text=black, font=\small, anchor=center] at ( 0.0, -1.25) {\text{115 $\mathrm \mu m$ FPE}}; 
            \node[text=black, font=\small, anchor=center] at ( 6.0, -1.25) {\text{550 $\mathrm \mu m$ FPE}}; 
        \end{tikzpicture}
    \end{subfigure}

    \begin{subfigure}[t]{1.00\textwidth}   
        \centering
        \vspace{-2mm}
        \caption{} 
        \begin{tikzpicture}[scale=1.00]
            \node[inner sep=0pt] (img) at (0,0) {\includegraphics[width=1\textwidth]{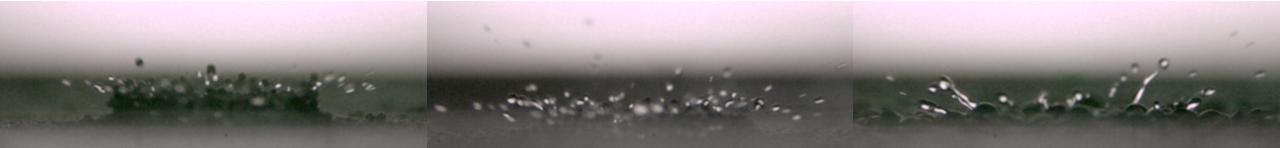}};

            \def\xval{7.70} 
            \def\yval{0.89}
            \draw[line width=0.5mm, draw=black, -] (\xval, \yval) -- (\xval+0.94, \yval); 
            \node[text=black, font=\footnotesize] at (\xval+0.47, \yval-0.19) {\textbf{2 mm}}; 

            \node[text=black, font=\small, anchor=center] at (-6.0, -1.25) {\text{231 $\mathrm \mu m$ G-FPE}};
            \node[text=black, font=\small, anchor=center] at ( 0.0, -1.25) {\text{327 $\mathrm \mu m$ G-PE}}; 
            \node[text=black, font=\small, anchor=center] at ( 6.0, -1.25) {\text{780 $\mathrm \mu m$ G-FPE}}; 
        \end{tikzpicture}
    \end{subfigure}

    \begin{subfigure}[t]{1.00\textwidth}   
        \centering
        \vspace{-2.5mm}
        \caption{}         
        \input{Figures/Fig2_Splash_Scaling}
    \end{subfigure}

    \caption{\textbf{
    Splashing dynamics on particle rafts. }
    \textbf{(a,b)} Impact outcomes mapped in the ($We$, $D_p$) plane for (a) clean water + untreated and (b) Glaco-treated (superhydrophobic) particle rafts, showing the transition from spreading to splashing across particle diameters and Weber numbers. The dashed lines show the empirical trend of the splashing threshold highlighting the progressive decrease in the critical Weber number as particle diameter increases. The Glaco treatment (superhydrophobic) produces a shallow particle immersion and large apparent contact angles that shift the splashing threshold to low Weber numbers.
    \textbf{(c,d)} Side-view snapshots of droplet impact at $t=0.05~\mathrm{ms}$, after impact at $We=346\pm3$, comparing splashing morphologies on untreated (c) and Glaco-treated (d) particle rafts. For both, the splashing morphology evolves from classical crown splashing on water to prompt splashing for small particles ($115 - 327~\mathrm{\mu m}$), before reverting to crown-like splashing for larger particles ($550$ and $780~\mathrm{\mu m}$).
    \textbf{(e)} Our scaling, $\Lambda$, collapse of all experiments in the ($We,~\Lambda$) plane. The green dashed line indicates the splashing threshold on clean water for reference.
    }  
    \label{figure2splash} 
    
\end{figure*} 

\subsection*{Splashing dynamics} 
As expected, splashing is influenced by both the raft characteristics and the impact properties. The influence of the raft surface coverage on splashing is first quantified by mapping impact outcomes across the Weber number and particle size (Fig.~\ref{figure2splash}a–b). For example, on a clean water (raft-free) surface,(crown) splash first occurs at $We\approx193$, this in agreement with previous studies\cite{motzkus2008study}. Under these conditions, the expanding lamella lifts off to form a well-defined crown that subsequently fragments into secondary droplets (splashing). In this regime, splashing is predominantly pool-driven, with the ejecta originating from the pool. In contrast, impacting a fully packed microparticle raft (as seen in Figure~\ref{figure1setup}c), fundamentally alters both the splashing threshold and the resulting dynamics. Unlike clean water, splashing on particle-coated interfaces becomes particle-size dependent where the ejecta volume can proceed from the pool or the impacting droplet (Fig.~\ref{figure2splash}c–d). For the smallest particles used here, i.e., $115\,\mathrm{\mu m}$ of untreated FPE (Fig.~\ref{figure2splash}a), splashing is hindered, with the onset occurring at $We\approx251$. This suggests that a densely packed monolayer of small particles stabilises the pool lamella and suppresses the growth of rim perturbations. In fact, when splashing does occur, it is predominantly prompt splashing, characterised by the ejection of $\sim10$ secondary microdroplets from the droplet rather than from the pool; a reduced crown height is also observed (Fig.~\ref{figure2splash}c). 

Intermediate FPE particle sizes ($231$ and $327\,\mathrm{\mu m}$) exhibit prompt splashing at lower Weber numbers than clean water ($We\approx163$), and increase the number of secondary microdroplets. As particle size increases across this range, prompt splashing becomes progressively more violent (i.e., more secondary droplets are generated) and the crown height is further diminished. For the largest untreated FPE particles ($550-780\,\mathrm{\mu m}$), splashing is triggered at even lower Weber numbers ($We\approx101$) and is characterised by a pronounced fingering of the crown lamella, which subsequently breaks up into secondary droplets (Fig.~\ref{figure2splash}c). Although the interface is not a rigid solid, a monolayer raft acts as a rough substrate, with its roughness controlled by the particle size. Under these conditions, a raft with particle diameter $D_p\gtrsim550\,\mathrm{\mu m}$ presents larger heterogeneity (and roughness) than a $D_p\lesssim327\,\mathrm{\mu m}$ raft. As with rough surfaces, large-amplitude roughness trigger rim perturbations that promote fingering and breakup\cite{iqbal2025droplet}. In contrast, smaller particles act as a quasi-continuous coating that damps interfacial deformation.

In fact, the splashing behaviour is not solely governed by particle size; both particle density and wettability determine the immersion depth and the contact angle which in turn control raft inertia and the lamella lift-off. For instance, untreated PE spheres ($\rho_P = 1.35\,\mathrm{g\,cm^{-3}}$) at $327$ and $780\,\mathrm{\mu m}$ consistently splash at lower Weber numbers than their lower-density FPE counterparts ($\rho_P=1.02\,\mathrm{g\,cm^{-3}}$). Conversely, hydrophilic glass spheres ($\rho_P= 2.50\,\mathrm{g\,cm^{-3}}$, $625\,\mathrm{\mu m}$) are substantially more submerged ($h\approx0.90$) and exhibit a small geometric contact angle ($\theta \approx37^\circ$), such that only a limited fraction of the particle protrudes above the interface. 
As a result, splashing occurs only at significantly higher Weber numbers ($We\approx160$), counter to expectations based on particle size and density alone. This demonstrates that the immersion depth plays a dominant role, alongside particle size and density, in setting the splashing threshold.

\definecolor{ElasticShade}{RGB}{190,215,245}
\definecolor{RigidShade}{RGB}{245,220,195}

\begin{figure*}[t!]
    \begin{subfigure}[t]{1.00\textwidth}   
        \centering
        \vspace{-7mm}
        \caption{} 
        \begin{tikzpicture}[scale=1.00]
            \node[inner sep=0pt] (img) at (0,0) {\includegraphics[width=1\textwidth]{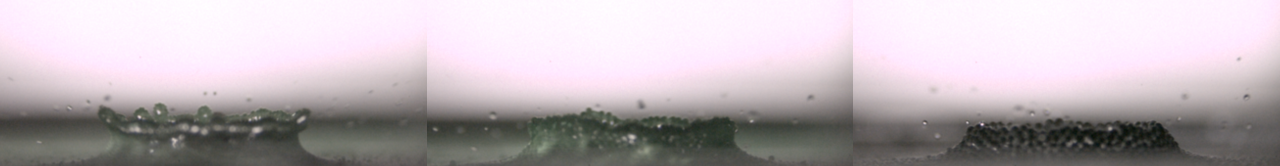}};

            \def\xval{7.90} 
            \def\yval{0.95}
            \draw[line width=0.5mm, draw=black, -] (\xval, \yval) -- (\xval+0.72, \yval); 
            \node[text=black, font=\footnotesize] at (\xval+0.36, \yval-0.19) {\textbf{2 mm}}; 
            
            \node[text=black, font=\small, anchor=center] at (-6.0, -1.40) {\text{$115~\mathrm{\mu m}$ FPE}}; 
            \node[text=black, font=\small, anchor=center] at ( 0.0, -1.40) {\text{$231~\mathrm{\mu m}$ FPE}}; 
            \node[text=black, font=\small, anchor=center] at ( 6.0, -1.40) {\text{$327~\mathrm{\mu m}$ PE}}; 
        \end{tikzpicture}
    \end{subfigure}
    \begin{subfigure}[t]{1.00\textwidth}   
        \centering
        \vspace{-2.5mm}
        \caption{} 
        \begin{tikzpicture}[scale=1.00]
            \node[inner sep=0pt] (img) at (0,0) {\includegraphics[width=1\textwidth]{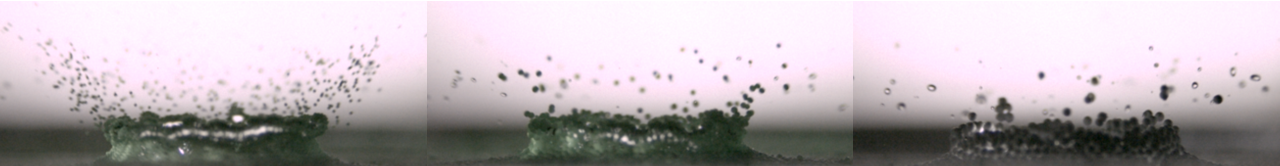}};
            
            \def\xval{7.90} 
            \def\yval{0.95}
            \draw[line width=0.5mm, draw=black, -] (\xval, \yval) -- (\xval+0.72, \yval); 
            \node[text=black, font=\footnotesize] at (\xval+0.36, \yval-0.19) {\textbf{2 mm}}; 

            \node[text=black, font=\small, anchor=center] at (-6.0, -1.40) {\text{$115~\mathrm{\mu m}$ G-FPE}}; 
            \node[text=black, font=\small, anchor=center] at ( 0.0, -1.40) {\text{$231~\mathrm{\mu m}$ G-FPE}}; 
            \node[text=black, font=\small, anchor=center] at ( 6.0, -1.40) {\text{$327~\mathrm{\mu m}$ G-PE}}; 
        \end{tikzpicture}
    \end{subfigure}

    \caption{\textbf{Impact-driven particle ejection during droplet splashing on FPE/PE rafts at $We\approx415$.}
    \textbf{(a,b)} Side-view snapshots at $t = 2.8~\mathrm{ms}$ after impact comparing particle splash for untreated (a) and Glaco-treated (b), illustrating the strong contrast between suppressed splash for hydrophobic particles and prolific radial ejection for superhydrophobic particles. For untreated rafts, only limited particle detachment is observed, whereas treated rafts exhibit widespread scattering due to minimal interfacial pinning.
    }  
    \label{figure3ParticleSplash} 
\end{figure*} 

In contrast, Glaco-treated (superhydrophobic) particles exhibit a systematic and monotonic dependence of the splashing threshold on particle size (Fig.~\ref{figure2splash}b). Their shallow immersion depths ($h<0.11$) and large apparent contact angles ($\theta>147^\circ$) effectively eliminate the wettability heterogeneities present in the untreated rafts, yielding a chemically homogeneous interface. As a result, particle size emerges as a primary parameter governing splashing.

The splashing threshold for small Glaco-treated particles, i.e., $115-327\,\mathrm{\mu m}$, is markedly reduced compared to untreated particles, decreasing from $We\approx251-163$ to $We\approx101-65$. This behaviour is also seen for large untreated and treated PE/FPE particles, i.e., $550-780\,\mathrm{\mu m}$, where the splashing threshold moves from $We\approx101$ to $We\approx65$. Notably, glass particles ($625\,\mathrm{\mu m}$), also follows this monotonic trend with the threshold reduced from $We\approx160$ to $We\approx65$. This behaviour reflects the combined effects of reduced immersion depth and enhanced effective roughness, as Glaco-treated particles sit high at the interface and produce pronounced geometric heterogeneities. Splashing in this regime experiences a large advancing dynamic contact angle, which is known to promote lamella lift-off and early breakup on superhydrophobic Glaco-trated solid surfaces \cite{quetzeri2019role}. Consistent with this past work, Glaco-treated rough solid surfaces exhibit similarly reduced splashing thresholds ($We\approx79$), reinforcing the role of the contact angle and interfacial roughness in facilitating splashing\cite{iqbal2025droplet}. As in the untreated case, the violence of the splash on superhydrophobic rafts increases with particle size, with larger particles producing more pronounced fingering and secondary droplet formation (Fig.~\ref{figure2splash}d).


Here, we scale these observations through a parameter that combines the influence of the three dominant factors on the impact dynamics, i.e. the raft roughness, particle inertia, and wettability. We propose a simple scaling argument that parametrises the splashing behaviour by taking into account the \emph{effective} geometric roughness ratio through the particle-size ratio $D_P/D$, the inertial contribution and interfacial anchoring of the particle raft through the density ratio $\rho_P/\rho$, and wettability through the contact angle. Under these conditions, the scaling is:
\begin{equation}
\Lambda = \left( \frac{D_P}{D} \right)
          \left( \frac{\rho_P}{\rho} \right)
          (1 - \cos\theta)^{3}
\end{equation}



Figure~\ref{figure2splash} shows our scaled experimental data; as seen the data collapses all the transitions from spreading to splashing onto a single behaviour in a ($We$, $\Lambda$) plane  across all materials and treatments (hydrophilic glass, untreated hydrophobic PE/FPE, and Glaco-treated superhydrophobic particles). Indeed, rafts with $\Lambda \gtrsim 0.5$ values (large particles, high density, or very hydrophobic surfaces) destabilise the droplet lamella and therefore splash is seen at comparatively low Weber numbers. In contrast, small or hydrophilic particles populate the opposite regime. The agreement of the data to this scaling demonstrates that splashing on particle rafts is not governed by any single property, but by a combined geometric–inertial–capillary resistance.

\definecolor{ElasticShade}{RGB}{190,215,245}
\definecolor{RigidShade}{RGB}{245,220,195}

\begin{figure*}[t!]
    \begin{subfigure}[t]{0.49\textwidth}   
        \centering
        \vspace{-7mm}
        \caption{}         
        \input{Figures/Fig4_Jet_untreated_Fr}
    \end{subfigure}
    \begin{subfigure}[t]{0.49\textwidth}   
        \centering
        \vspace{-7mm}
        \caption{}         
        \pgfplotsset{compat=1.17} 

\pgfdeclareplotmark{hexagon*}{%
  \pgfpathmoveto{\pgfqpoint{1\pgfplotmarksize}{0pt}}%
  \pgfpathlineto{\pgfqpoint{0.5\pgfplotmarksize}{0.8660254\pgfplotmarksize}}%
  \pgfpathlineto{\pgfqpoint{-0.5\pgfplotmarksize}{0.8660254\pgfplotmarksize}}%
  \pgfpathlineto{\pgfqpoint{-1\pgfplotmarksize}{0pt}}%
  \pgfpathlineto{\pgfqpoint{-0.5\pgfplotmarksize}{-0.8660254\pgfplotmarksize}}%
  \pgfpathlineto{\pgfqpoint{0.5\pgfplotmarksize}{-0.8660254\pgfplotmarksize}}%
  \pgfpathclose
  \pgfusepathqfillstroke
}

\definecolor{Green}{RGB}{50,155,105}    
\definecolor{Pink}{RGB}{213,81,156}     
\definecolor{Blue}{RGB}{54,80,163}      
\definecolor{Yellow}{RGB}{244,196,48}   
\definecolor{Orange}{RGB}{245,127,34}   
\definecolor{Purple}{RGB}{142, 68, 173} 
\definecolor{Red}{RGB}{150,30,60}       

\definecolor{Cyan}{RGB}{32,170,170}     
\definecolor{Grey}{RGB}{120,120,120}    

\definecolor{LightPurple}{RGB}{204,204,255}  

\begin{tikzpicture}
\begin{axis}[
width=1.55\figW,
height=0.975\figH,
at={(0\figW,0\figH)},
    xlabel={\small $Fr$},
    xmin=50, xmax=500,
    ymin=0, ymax=8,
    xtick={50, 150, 250, 350, 450}, 
    minor x tick num=1,  
    minor y tick num=1,  
    yticklabel style={/pgf/number format/.cd, fixed, fixed zerofill, precision=0},
    ticklabel style = {font=\footnotesize}
    ]

\definecolor{ElasticShade}{RGB}{190,215,245}
\definecolor{RigidShade}{RGB}{245,220,195}

\definecolor{ElasticDark}{RGB}{120,150,150}   
\definecolor{RigidDark}{RGB}{160,140,110}     






\addplot[domain=70:480, samples=2, color=Green, dashed, line width=0.5pt]
{0.0153*x + 0.4018};


\addplot[only marks, color=Blue, mark=pentagon*, mark options={solid, fill=white}, mark size=1.5pt, line width=0.5pt]
table[row sep=crcr]{
471.3	5.68	\\
410.4	5.26	\\
348.9	5.01	\\
285.3	4.32	\\
216.6	3.90	\\
181.3	3.14	\\
145.4	2.45	\\
114.0	1.84	\\
74.1	1.09	\\
};

\addplot[domain=70:480, samples=2, color=Blue, dotted, line width=0.8pt]
{0.0113*x + 0.8159};


\addplot[only marks, color=Purple, mark=pentagon*, mark options={solid, fill=white}, mark size=1.5pt, line width=0.5pt]
table[row sep=crcr]{
476.4	7.26	\\
415.1	6.60	\\
353.7	7.09	\\
286.0	5.23	\\
218.9	4.47	\\
182.8	3.86	\\
148.0	2.39	\\
116.0	1.26	\\
74.0	0.88	\\
};

\addplot[domain=70:480, samples=2, color=Purple, dotted, line width=0.8pt]
{0.0167*x + 0.1119};


\addplot[only marks, color=Red, mark=pentagon*, mark options={solid, fill=white}, mark size=1.5pt, line width=0.5pt]
table[row sep=crcr]{
479.1	6.39	\\
416.4	6.13	\\
352.7	6.17	\\
286.7	5.29	\\
220.5	4.55	\\
185.0	3.49	\\
149.7	2.79	\\
114.0	1.90	\\
74.3	0.50	\\
};

\addplot[domain=70:480, samples=2, color=Red, dotted, line width=0.8pt]
{0.0141*x + 0.5664};


\addplot[only marks, color=Yellow, mark=pentagon*, mark options={solid, fill=white}, mark size=1.5pt, line width=0.5pt]
table[row sep=crcr]{
476.4	3.55	\\
415.1	3.53	\\
353.7	2.82	\\
286.0	1.95	\\
218.9	1.73	\\
182.8	1.30	\\
148.0	0.85	\\
116.0	0.64	\\
};

\addplot[domain=110:480, samples=2, color=Yellow, dotted, line width=0.8pt]
{0.0086*x - 0.3245};


\addplot[only marks, color=Orange, mark=pentagon*,  mark options={solid, fill=white}, mark size=1.5pt, line width=0.5pt]
table[row sep=crcr]{
479.1	3.01	\\
416.4	2.44	\\
352.7	2.47	\\
286.7	1.40	\\
220.5	0.95	\\
185.0	1.09	\\
149.7	0.70	\\
};

\addplot[domain=145:480, samples=2, color=Orange, dotted, line width=0.8pt]
{0.0071*x - 0.4055};


\addplot[only marks, color=Pink, mark=square*,   mark options={solid, fill=white}, mark size=1.5pt, line width=0.5pt]
table[row sep=crcr]{
471.3	2.46	\\
410.4	2.35	\\
348.9	1.74	\\
285.3	1.11	\\
216.6	0.85	\\
181.3	0.83	\\
145.4	0.83	\\
};

\addplot[domain=140:480, samples=2, color=Pink, dotted, line width=0.8pt]
{0.0057*x-0.2358};


\addplot[only marks, color=Cyan, mark=triangle*, mark options={solid, fill=white}, mark size=1.5pt, line width=0.5pt]
table[row sep=crcr]{
479.1	6.43	\\
416.4	6.15	\\
352.7	5.85	\\
286.7	5.12	\\
220.5	3.86	\\
185.0	3.25	\\
149.7	2.35	\\
114.0	1.93	\\
74.3	0.64	\\
};
\addplot[domain=70:480, samples=2, color=Cyan, dotted, line width=0.8pt]
{0.0143*x+0.3325};





\addplot[domain=70:480, samples=2, color=Green, dashed, line width=0.5pt]
{0.0153*x + 0.4018};

\addplot[domain=110:480, samples=2, color=Blue, dashed, line width=0.5pt]
{0.0114*x + 0.5065};

\addplot[domain=110:480, samples=2, color=Purple, dashed, line width=0.5pt]
{0.0118*x + 0.118};

\addplot[domain=110:480, samples=2, color=Red, dashed, line width=0.5pt]
{0.0127*x - 0.6621};

\addplot[domain=140:480, samples=2, color=Yellow, dashed, line width=0.5pt]
{0.0089*x - 0.7192};

\addplot[domain=180:480, samples=2, color=Orange, dashed, line width=0.5pt]
{0.0061*x - 0.4928};

\addplot[domain=110:480, samples=2, color=Pink, dashed, line width=0.5pt]
{0.0106*x-0.733};

\addplot[domain=110:480, samples=2, color=Cyan, dashed, line width=0.5pt]
{0.0118*x-0.1052};


\end{axis}
\end{tikzpicture}
    \end{subfigure}

        \begin{subfigure}[t]{1.00\textwidth}   
        \centering
        \vspace{0mm}
        \caption{} 
        \begin{tikzpicture}[scale=1.00]
            \node[inner sep=0pt] (img) at (0,0) {\includegraphics[width=1\textwidth]{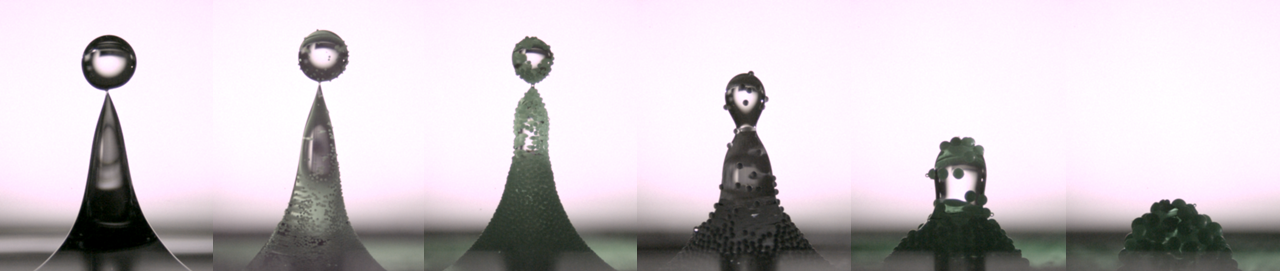}};
            
            \def\xval{8.10} 
            \def\yval{1.70}
            \draw[line width=0.5mm, draw=black, -] (\xval, \yval) -- (\xval+0.51, \yval); 
            \node[text=black, font=\footnotesize] at (\xval+0.255, \yval-0.19) {\textbf{2 mm}}; 
            
            \node[text=black, font=\small, anchor=center] at (-7.4, -2.10) {\text{Water}}; 
            \node[text=black, font=\small, anchor=center] at (-4.5, -2.10) {\text{$115~\mathrm \mu m$ FPE}}; 
            \node[text=black, font=\small, anchor=center] at (-1.5, -2.10) {\text{$231~\mathrm \mu m$ FPE}}; 
            \node[text=black, font=\small, anchor=center] at ( 1.5, -2.10) {\text{$327~\mathrm \mu m$ PE}}; 
            \node[text=black, font=\small, anchor=center] at ( 4.5, -2.10) {\text{$550~\mathrm \mu m$ FPE}}; 
            \node[text=black, font=\small, anchor=center] at ( 7.4, -2.10) {\text{$780~\mathrm \mu m$ FPE}}; 
            
        \end{tikzpicture}
    \end{subfigure}
    \begin{subfigure}[t]{1.00\textwidth}   
        \centering
        \vspace{-3mm}
        \caption{} 
        \begin{tikzpicture}[scale=1.00]
            \node[inner sep=0pt] (img) at (0,0) {\includegraphics[width=1\textwidth]{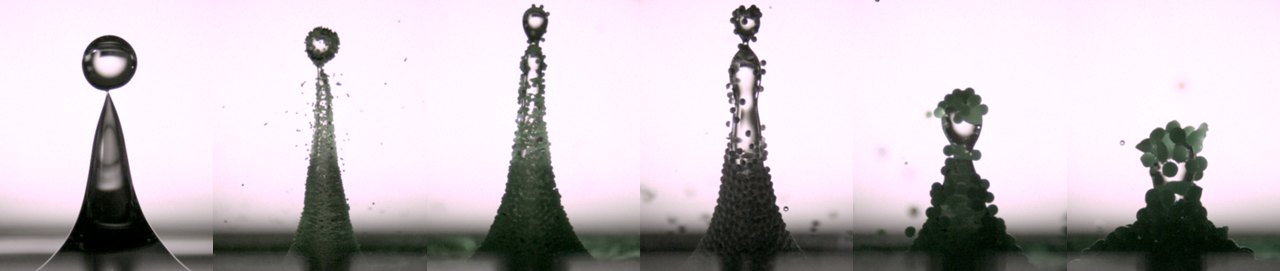}};

            \def\xval{8.10} 
            \def\yval{1.70}
            \draw[line width=0.5mm, draw=black, -] (\xval, \yval) -- (\xval+0.51, \yval); 
            \node[text=black, font=\footnotesize] at (\xval+0.255, \yval-0.19) {\textbf{2 mm}}; 
            
            \node[text=black, font=\small, anchor=center] at (-7.4, -2.10) {\text{Water}}; 
            \node[text=black, font=\small, anchor=center] at (-4.5, -2.10) {\text{115 $\mathrm \mu m$ G-FPE}}; 
            \node[text=black, font=\small, anchor=center] at (-1.5, -2.10) {\text{231 $\mathrm \mu m$ G-FPE}}; 
            \node[text=black, font=\small, anchor=center] at ( 1.5, -2.10) {\text{$327~\mathrm \mu m$ G-PE}}; 
            \node[text=black, font=\small, anchor=center] at ( 4.5, -2.10) {\text{550 $\mathrm \mu m$ G-FPE}}; 
            \node[text=black, font=\small, anchor=center] at ( 7.4, -2.10) {\text{780 $\mathrm \mu m$ G-FPE}}; 
            
        \end{tikzpicture}
    \end{subfigure}
    
    \caption{\textbf{Influence of the particle raft size on the Worthington jet height. Solid symbols represent untreated particles, while hollow symbols and represent Glaco-treated particles; dashed and dotted lines added to guide the eye.}
    \textbf{(a,b)} Maximum dimensionless jet height, $H_J/D$, as a function of the Froude number for clean water and particle rafts of varying diameter; (a) shows results for untreated particles and (b) shows Glaco-treated particles. For untreated rafts, the onset of jetting shifts systematically to higher $Fr$ with increasing particle size with the maximum jet height decreasing monotonically. In contrast, Glaco-treated rafts exhibit earlier jet formation and a non-monotonic dependence of jet height on particle diameter, with the tallest jets observed for intermediate sizes ($231$ and $327~\mathrm{\mu m}$). In panel (b), dashed lines for untreated rafts are shown for reference.
    \textbf{(c,d)} Side-view snapshots of the Worthington jet at the instant of maximum height at $Fr=417\pm3$, comparing untreated (c) and Glaco-treated (d) particle rafts with clean water.
    }  
    \label{figure4jet} 
\end{figure*}



\subsection*{Impact-driven particle ejection during splashing}

Beyond controlling the splashing threshold and behaviour, wettability also governs whether splashing is accompanied by particle ejection at the moment of impact. Particle ejection occurs during the early stages of droplet–raft interaction, as horizontal momentum from the impacting droplet is transferred to the particle monolayer during the lamella expansion and cavity formation. For untreated particle rafts, impact-driven particle ejection is strongly suppressed across the entire parameter space explored. Small untreated particles ($115-231\,\mathrm{\mu m}$) exhibit only limited ejection, and only at the highest impact speeds ($We \gtrsim 277$), typically expelling fewer than 5–10 particles (Fig.~\ref{figure3ParticleSplash}a). Their substantial immersion depths ($h \approx 0.4-0.9$) result in strong capillary adhesion to the interface\cite{Binks2006colloidal}, such that significant impulse is required to detach particles from the raft. For larger untreated particles ($\ge327\,\mathrm{\mu m}$), impact-driven particle splash is never observed. In this regime, increased particle inertia and larger interfacial contact areas further inhibit displacement, and splashing proceeds almost exclusively through liquid fragmentation, with particles remaining bound to the interface.

In stark contrast, Glaco-treated superhydrophobic particles exhibit enhanced impact-driven particle ejection. For small treated particles ($115-327\,\mathrm{\mu m}$), particle splash initiates at low Weber numbers ($We\approx65$) and intensifies rapidly with impact energy, ejecting tens to hundreds of particles radially from the impact site (Fig.~\ref{figure3ParticleSplash}b). Their extremely shallow immersion depths ($h \approx 0.04-0.11$) render the particles quasi-buoyant, resting on the interface with minimal capillary pinning\cite{Binks2006colloidal, kralchevsky2001particles}. As a result, the horizontal impulse associated with lamella expansion readily overcomes weak inter-particle cohesion, leading to rapid rupture and dispersal of the raft.

For larger Glaco-treated particles ($550-780\,\mathrm{\mu m}$), impact-driven particle ejection is only observed at the highest Weber numbers. At lower impact inertia, splashing remains predominantly liquid-dominated, with particles remaining bound to the interface despite vigorous lamella breakup. This transition reflects the increased impulse required to accelerate the larger particle mass. The suppression of impact-driven ejection is most pronounced for glass spheres, which are never observed to eject during splashing, irrespective of surface treatment. Their high density ($\rho_P\approx2.50\,\mathrm{g\,cm^{-3}}$) makes particle inertia dominant, pinning them to the interface during impact and preventing detachment even under superhydrophobic conditions.

Together, these observations demonstrate that impact-driven particle ejection during splashing is governed by a competition between capillary adhesion and particle inertia. While superhydrophobic treatment removes wettability-induced pinning and promotes particle mobility, particle size and density ultimately determine whether splashing transitions from a purely liquid-dominated process to one accompanied by direct particle entrainment.

\subsection*{Jet dynamics} \label{Sec:Jet}
Our results indicate that raft particles modify the dynamics of the cavity depth and collapse in Worthington jets. Here, based in our observations, we propose a scaling law for the Worthington jet height and particle entrainment. In our analysis, the maximum dimensionless Worthington jet height, $H_J/D$, is measured as a function of the Froude number (Fig.~\ref{figure4jet}a–b). Following prior work, we define a characteristic Worthington jet height as the jet length at the point where it exceeds half of the drop diameter, $H_J/D>0.5$. This characteristic length thereby excludes weak surface instabilities that do not develop into a coherent vertical (Worthington) jet. In our experiments, on a clean water surface, the onset for Worthington jet formation is found at $Fr\gtrsim70$, which is consistent with earlier works \cite{michon2017jet}.  

Results indicate that untreated particle rafts delay and suppres jet formation, as seen in Figure~\ref{figure4jet}a. For untreated PE and FPE rafts, the onset of jet formation systematically shifts to higher $Fr$ numbers with increasing particle size. For example, jets appear for $Fr \gtrsim 110$ for particle diameters of $115$, $231$ and $327\,\mathrm{\mu m}$, while larger particles require substantially faster impacts, with thresholds of $Fr \gtrsim 148$ for $550\,\mathrm{\mu m}$ and $Fr \gtrsim 184$ for $780\,\mathrm{\mu m}$. Beyond the onset, the maximum jet height decreases monotonically with particle size, indicating that large particles impede the upward focusing of momentum required for cavity collapse (Fig.~\ref{figure4jet}c). In fact, changing particle density, at fixed size, has only a weak influence on the jet height for untreated rafts, i.e. PE particles ($\rho_P = 1.35\,\mathrm{g\,cm^{-3}}$) and FPE particles ($\rho_P = 1.02\,\mathrm{g\,cm^{-3}}$) of comparable diameters reach similar jet heights, suggesting that particle inertia alone does not control the jet suppression. A notable exception is seen by hydrophilic glass spheres ($D_P = 625\,\mathrm{\mu m}$), for which jetting occurs at $Fr \gtrsim 110$, comparable to much smaller PE particles. High-speed imaging shows that these glass particles submerge and are readily displaced downward upon impact, leaving a clear imprint of the crater on the pool surface. As a result, the collapsing cavity encounters a locally depleted interfacial resistance, partially restoring the momentum focusing required for jet formation despite the larger particle size and density. A qualitatively different behaviour emerges for Glaco-treated, superhydrophobic particle rafts (Fig.~\ref{figure4jet}b). In all Glaco-treated cases, jet formation occurs at lower Froude numbers than for their size-matched untreated particles (except G-Glass), indicating early and efficient momentum focusing. For G-PE and G-FPE particle diameters of $115$, $231$ and $327\,\mathrm{\mu m}$, jets appear at $Fr\gtrsim70$, matching clean water. Larger Glaco-treated particles exhibit delayed jetting relative to smaller ones, but still at substantially lower thresholds than untreated rafts, with onsets at $Fr \gtrsim 110$ for $550\,\mathrm{\mu m}$ and $Fr \gtrsim 145$ for $780\,\mathrm{\mu m}$. This indicates that, across $Fr$, superhydrophobic treatment makes the interface easier to mobilise and reorganise during collapse, thereby restoring earlier jet formation.

\definecolor{ElasticShade}{RGB}{190,215,245}
\definecolor{RigidShade}{RGB}{245,220,195}

\usepgfplotslibrary{fillbetween}

\definecolor{SharpWShade}{RGB}{33,33,33}     
\definecolor{SoftWShade}{RGB}{117,107,177}   
\definecolor{FlatVShade}{RGB}{255,237,160}   
\definecolor{SharpVShade}{RGB}{158,202,225}  
\definecolor{SoftVShade}{RGB}{33,113,181}    

\begin{figure*}[t!]
    \begin{subfigure}[t]{0.49\textwidth}   
        \centering
        \vspace{-7mm}
        \caption{}         
        \input{Figures/Fig5_CavityDepth_Untreated}
    \end{subfigure}
    \begin{subfigure}[t]{0.49\textwidth}   
        \centering
        \vspace{-7mm}
        \caption{}         
        \input{Figures/Fig5_CavityDepth_Treated}
    \end{subfigure}

    \begin{subfigure}[t]{1.00\textwidth}   
        \centering
        \vspace{0mm}
        \caption{} 
        \begin{tikzpicture}[scale=1.00]
            \node[inner sep=0pt] (img) at (0,0) {\includegraphics[width=1\textwidth]{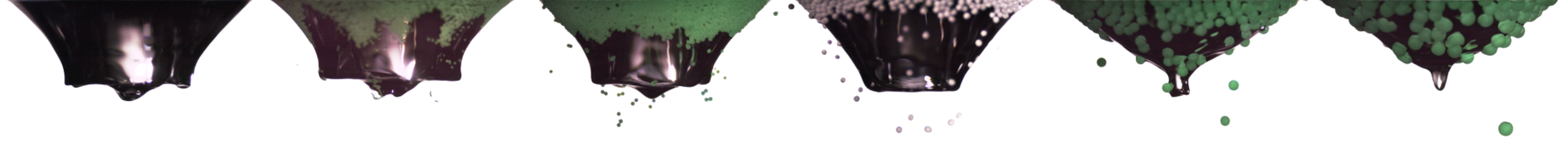}};

            \draw[SharpWShade, line width=2.5pt, opacity = 0.5] ( -8.70, -0.75) rectangle ( -0.047, 0.85);  
            \draw[FlatVShade,  line width=2.5pt, opacity = 0.5] ( 0.038, -0.75) rectangle (  2.86, 0.85);
            \draw[SharpVShade, line width=2.5pt, opacity = 0.5] ( 2.945, -0.75) rectangle (  8.76, 0.85);   
            
            \node[text=black, font=\small, anchor=center] at (-7.4, -1.00) {\text{Water}}; 
            \node[text=black, font=\small, anchor=center] at (-4.5, -1.00) {\text{$115~\mathrm \mu m$ FPE}}; 
            \node[text=black, font=\small, anchor=center] at (-1.5, -1.00) {\text{$231~\mathrm \mu m$ FPE}}; 
            \node[text=black, font=\small, anchor=center] at ( 1.5, -1.00) {\text{$327~\mathrm \mu m$ PE}}; 
            \node[text=black, font=\small, anchor=center] at ( 4.5, -1.00) {\text{$550~\mathrm \mu m$ FPE}}; 
            \node[text=black, font=\small, anchor=center] at ( 7.4, -1.00) {\text{$780~\mathrm \mu m$ FPE}}; 
            
        \end{tikzpicture}
    \end{subfigure}
    \begin{subfigure}[t]{1.00\textwidth}   
        \centering
        \vspace{-3mm}
        \caption{} 
        \begin{tikzpicture}[scale=1.00]
            \node[inner sep=0pt] (img) at (0,0) {\includegraphics[width=1\textwidth]{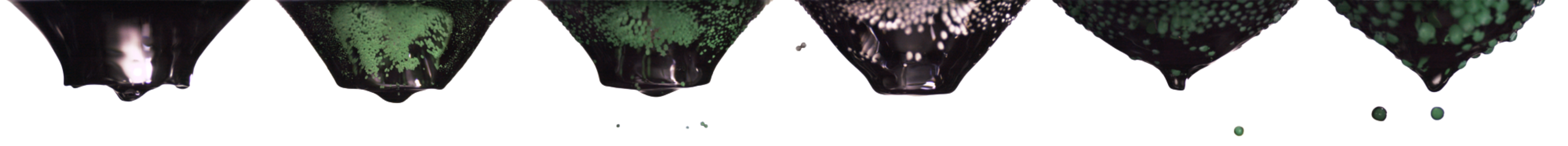}};
            
            \def\xval{8.15} 
            \def\yval{-0.60}
            \draw[line width=0.5mm, draw=black, -] (\xval, \yval) -- (\xval+0.39, \yval); 
            \node[text=black, font=\footnotesize] at (\xval+0.18, \yval+0.19) {\textbf{2 mm}}; 

            \draw[SharpWShade, line width=2.5pt, opacity = 0.5] ( -8.70, -0.75) rectangle ( -5.885, 0.85);  
            \draw[SoftWShade,  line width=2.5pt, opacity = 0.5] ( -5.80, -0.75) rectangle ( -0.030, 0.85);  
            \draw[FlatVShade,  line width=2.5pt, opacity = 0.5] ( 0.055, -0.75) rectangle (  2.85, 0.85);
            \draw[SharpVShade, line width=2.5pt, opacity = 0.5] ( 2.935, -0.75) rectangle (  8.76, 0.85);   
            
            \node[text=black, font=\small, anchor=center] at (-7.4, -1.00) {\text{Water}}; 
            \node[text=black, font=\small, anchor=center] at (-4.5, -1.00) {\text{115 $\mathrm \mu m$ G-FPE}}; 
            \node[text=black, font=\small, anchor=center] at (-1.5, -1.00) {\text{231 $\mathrm \mu m$ G-FPE}}; 
            \node[text=black, font=\small, anchor=center] at ( 1.5, -1.00) {\text{$327~\mathrm \mu m$ G-PE}}; 
            \node[text=black, font=\small, anchor=center] at ( 4.5, -1.00) {\text{550 $\mathrm \mu m$ G-FPE}}; 
            \node[text=black, font=\small, anchor=center] at ( 7.4, -1.00) {\text{780 $\mathrm \mu m$ G-FPE}}; 
            
        \end{tikzpicture}
    \end{subfigure}
    \caption{\textbf{Cavity depth and collapse morphology underlying jet formation on particle rafts. Solid symbols and dashed lines represent untreated particles, while hollow symbols and dotted lines represent Glaco-treated particles.}
    \textbf{(a,b)} Maximum dimensionless cavity depth, $H_C/D$, as a function of the Froude number $Fr$ for clean water and particle rafts composed of FPE and PE particles of varying diameter. Untreated particles seen in (a) and Glaco-treated seen in (b). For both conditions, cavity depth decreases monotonically with increasing particle size. In panel (b), dashed lines for untreated rafts are shown for reference. Shaded backgrounds denote the dominant cavity collapse shape identified from side-view imaging (legend). 
    \textbf{(c,d)} Representative side-view images of cavity collapse for untreated (c) and Glaco-treated (d) particle rafts at $Fr\approx471$, illustrating the transition between sharp W, soft W, flat V and sharp V morphologies with increasing particle size and surface treatment. Shaded boundaries denote the dominant cavity collapse shape.
    }  
    \label{figure5cavity} 
\end{figure*} 

In contrast to untreated rafts, Glaco-treated surfaces show a non-monotonic variation of jet height with particle size. The tallest jets occur for intermediate diameters ($231$ and $327\,\mathrm{\mu m}$), reaching heights comparable to, or marginally exceeding, those on clean water (Fig.~\ref{figure4jet}d). Although smaller particles ($115\,\mathrm{\mu m}$) generate lower jets, their heights remain systematically higher than in the size-matched untreated case. A similar enhancement is observed for larger Glaco-treated particles ($550$ and $780\,\mathrm{\mu m}$), which also produce taller jets than their untreated counterparts. 
The sole exception to this behaviour is provided by Glaco-treated glass spheres, where jet formation is delayed to $Fr\gtrsim145$, and the maximum jet height is markedly lower than for glass particles. Unlike in the untreated case, where hydrophilic glass spheres submerge and are expelled from the interface during impact. The superhydrophobic coating prevents submergence, forcing the dense particles to be accelerated upward together with the rising jet. This added inertial load limits vertical momentum focusing and suppresses jet growth. Once wettability effects are removed, particle density becomes a controlling factor: although Glaco-treated polyethylene (G-FPE $\&$ G-PE) particles of different densities ($1.02$ \& $1.35\,\mathrm{g\,cm^{-3}}$ respectively) produce comparable jet heights at fixed particle size, the substantially denser ($2.50\,\mathrm{g\,cm^{-3}}$) glass particles yield the lowest jet heights. These results demonstrate that particle size, wettability, and inertia control jet formation and height, motivating a closer examination of the effect of upstream cavity formation and collapse dynamics on the jet behaviour.

\subsection*{Cavity dynamics underlying jet formation} \label{Sec:Cavity}
Our experiments demonstrate that regarless of the particle wettability, the Worthington cavity depth decreases monotonically with increasing particle size. Figures~\ref{figure5cavity}a-b reports the maximum cavity depth, $H_C/D$, measured relative to the undisturbed free surface, for both untreated and Glaco-treated particle rafts, respectively. As observed, on a clean water surface, the impact produces the deepest cavities across the explored Froude-number range. This indicates an efficient downward momentum transfer and unobstructed cavity growth. Introducing a particle raft systematically reduces the cavity depth, with a clear decrease as particle size increases. Cavity depth decreases progressively from $115$ to $780\,\mathrm{\mu m}$ particles, for both untreated and Glaco-treated rafts, indicating that larger particles increasingly impede cavity formation irrespective of surface treatment. Unlike the jet height, the cavity depth exhibits monotonic dependence on particle size for Glaco-treated rafts, with $H_C/D$ decreasing with increasing particle diameter. We note that a systematic difference between treated and untreated rafts emerges as particle size increases (Fig.~\ref{figure5cavity}b). For the smallest particles ($115$ and $231\,\mathrm{\mu m}$), the cavity depths for both rafts are nearly indistinguishable, indicating that surface treatment has little influence on the initial cavity penetration. However, this difference grows steadily with particle diameter: Glaco-treated rafts produce deeper cavities than their untreated counterparts for $327\,\mathrm{\mu m}$ particles, with the disparity becoming more pronounced for $550$ and $780\,\mathrm{\mu m}$ particles. Taken together, these observations indicate that while deeper cavity formation explains the enhanced jetting on Glaco-treated rafts for large particles, it does not resolve the relatively stronger jet enhancement observed at smaller particle sizes. Consequently, the governing mechanism must instead lie in the cavity collapse. In the following section, we examine the cavity shapes and collapse dynamics in detail to establish how particle wettability and size control capillary-wave focusing and vertical momentum concentration during Worthington jet formation.

\definecolor{ElasticShade}{RGB}{190,215,245}
\definecolor{RigidShade}{RGB}{245,220,195}

\begin{figure*}[t!]
    \begin{subfigure}[t]{1.00\textwidth}   
        \centering
        \vspace{0mm}
        \begin{tikzpicture}[scale=1.00]
            \node[inner sep=0pt] (img) at (0,0) {\includegraphics[width=1\textwidth]{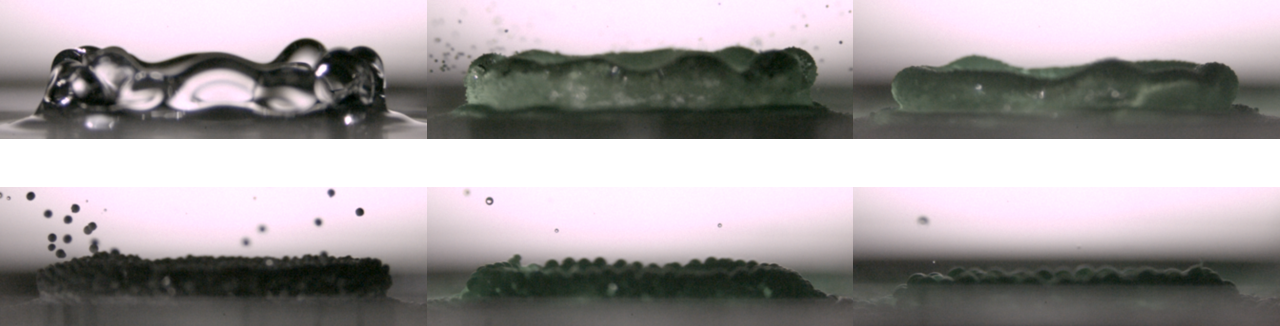}};
            
            \def\xval{7.90} 
            \def\yval{2.10}
            \draw[line width=0.5mm, draw=black, -] (\xval, \yval) -- (\xval+0.7, \yval); 
            \node[text=black, font=\footnotesize] at (\xval+0.35, \yval-0.19) {\textbf{2 mm}}; 
            
            \node[text=black, font=\small, anchor=center] at (-6.0,  0.1) {\text{Water}};
            \node[text=black, font=\small, anchor=center] at ( 0.0,  0.1) {\text{$115~\mathrm{\mu m}$ G-FPE}};
            \node[text=black, font=\small, anchor=center] at ( 6.0,  0.1) {\text{$231~\mathrm{\mu m}$ FPE}}; 
            \node[text=black, font=\small, anchor=center] at (-6.0, -2.5) {\text{$327~\mathrm{\mu m}$ G-PE}};
            \node[text=black, font=\small, anchor=center] at ( 0.0, -2.5) {\text{$550~\mathrm{\mu m}$ FPE}};
            \node[text=black, font=\small, anchor=center] at ( 6.0, -2.5) {\text{$780~\mathrm{\mu m}$ G-FPE}}; 
        \end{tikzpicture}
    \end{subfigure}
    \caption{\textbf{Wave swell suppression by particle rafts of increasing particle size.}
    Side-view snapshots at $t=8.7~\mathrm{ms}$ after impact for a clean water surface and particle rafts at $Fr\approx471$. The wave swell amplitude decreases systematically with increasing particle size, reflecting the increasing inertial resistance of the particle layer.
    }  
    \label{figure6WaveSwell} 
\end{figure*}
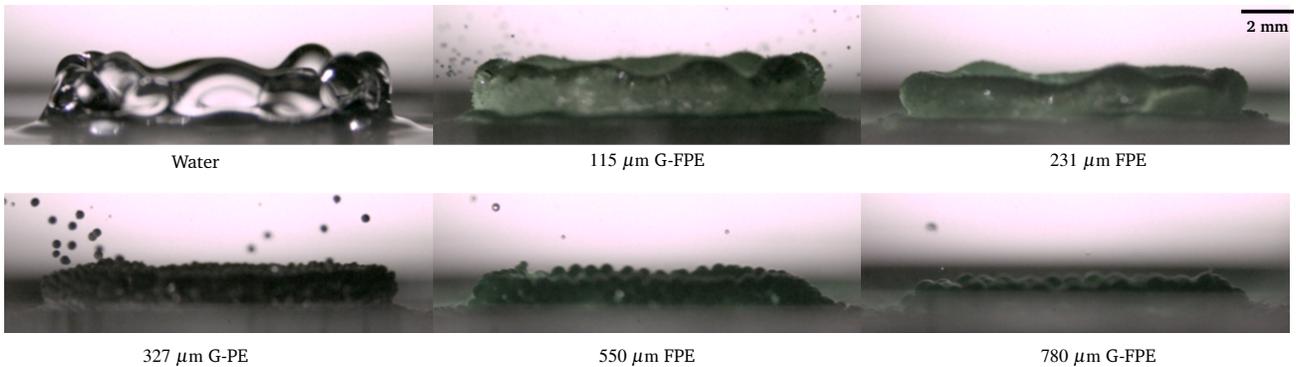 

\subsection*{Cavity collapse modes}\label{Sec:CavityModes} 

Figures \ref{figure5cavity}c-d present cavity shapes and collapse sequences. As observed, on a clean water surface, the cavity dynamics following droplet impact exhibit a systematic progression with increasing $Fr$, governed by the interplay between the inertia of the collapsing wave swell and capillary retraction. This sequence follows established regimes for droplet impact on deep pools \cite{ray2015regimes, michon2017jet, das2022evolution}. At high impact speeds ($Fr>300$), a pronounced wave swell forms at the cavity rim with its inward collapse generates a strong capillary-inertia wave that travels down the cavity wall. This wave induces a local re-expansion at the cavity floor, pinching it into a sharp W-shape. The subsequent collapse focuses momentum into a thick, slow Worthington jet that pinches off a single large droplet. At moderate $Fr$, the wave swell and the resulting capillary wave weaken. The cavity floor does not re-expand markedly, resulting in a soft W-shape or, at lower $Fr$, a sharp V-shape characteristic of a more capillary-driven collapse. The retraction of these shapes produces a faster, thinner jet susceptible to capillary instabilities, which breaks into multiple smaller droplets. This morphological transition from a sharp W to a sharp V with decreasing $Fr$ directly dictates the jet's morphology and the resulting droplet-size distribution\cite{ray2015regimes, michon2017jet, das2022evolution}.

The introduction of a particle raft fundamentally alters the cavity collapse sequence observed for a clean water surface. The raft’s properties, particle size and, critically, wettability, dictate whether the cavity follows a water-like collapse or a new, particle-mediated pathway. The particles control the efficiency of vertical momentum focusing and the characteristics of the ejected jet and droplets. For untreated particle rafts, the cavity morphology undergoes a systematic transition as particle size increases, shifting the collapse regime away from that of clean water. For the smallest particles ($115$ and $231\,\mathrm{\mu m}$), the cavity shape at high $Fr$ ($\gtrsim300$) remains a sharp W, similar to clean water. This indicates that a pronounced wave swell is still able to form and collapse, as visualised in Fig.~\ref{figure6WaveSwell}, to generate the capillary–inertia wave that induces base re-expansion. The resulting jet is consequently slow and thick, pinching off a single large droplet. However, as the particle size increases to $327\,\mathrm{\mu m}$, this capacity is disrupted. The cavity at high $Fr$ forms only a soft W (or a flat V) signalling a weakened wave swell. By $550$ and $780\,\mathrm{\mu m}$ particle size, the cavity is exclusively sharp V-shaped even at the highest impact energies ($Fr=471$). This morphological shift (from W to V with increasing particle size) demonstrates that larger, more massive rafts progressively suppress the formation and inward collapse of the wave swell, consistent with the systematic reduction in swell amplitude observed across particle sizes. The raft’s inertia and frictional dissipation impede the radial motion necessary to build up the wave swell, preventing the local re-expansion at the cavity floor. Consequently, the collapse mechanism transitions from the inertia-capillary wave-mediated process (sharp W) to a more direct, capillary-driven closure (sharp V). For the largest particles ($550, 780\,\mathrm{\mu m}$), the jet is slow and thick despite the cavity being a sharp V. This apparent contradiction is resolved by the role of particle entrainment. High-speed imaging confirms that these large particles are carried upward within the jet (Fig.~\ref{figure4jet}c-d). This particle-laden jet is more stable against capillary breakup and remains uniform and thick presenting a slower ascent and reduced height. The raft thus imposes a dual suppression: it first inhibits the W-forming collapse, and then the entrained particles dampen the jet dynamics.

Superhydrophobic treatment of the particle raft instigates a qualitatively different cavity evolution, characterised by the early ejection of particles and a consequent change in the energy partitioning during collapse. For small Glaco-treated particles ($115$ and $231\,\mathrm{\mu m}$), the cavity at high $Fr$ forms a soft W, distinct from the sharp W of clean water or untreated rafts of the same size. This is a pivotal observation: while cavity depth remains comparable to untreated cases, the shape is fundamentally different. 
The soft W and flat V indicate a diminished wave swell. Since the superhydrophobic coating enables immediate particle ejection during the initial splash and cavity opening, a significant portion of the impact energy is thereby diverted into kinetic energy of scattered particles rather than into the potential energy of the wave swell. With a less massive and pronounced swell, the ensuing capillary-inertia wave is weaker, resulting in a softer re-expansion at the cavity floor (soft W or flat V versus sharp W). This altered collapse has profound consequences for the jet. The softer, less focused collapse does not produce the slow, thick Rayleigh jet. Instead, it facilitates a more efficient capillary retraction of the cavity walls, sparking a high-speed and thin jet. This explains the non-monotonic jet height data: for $231$ and $327\,\mathrm{\mu m}$ Glaco-treated particles, the cavity depth is similar to that of the untreated ones, but the change in shape (soft W/flat V) leads to a faster, taller jet.
As with untreated rafts, increasing the size of Glaco-treated particles eventually overwhelms this effect. For larger particles ($550$ and $780\,\mathrm{\mu m}$), the cavity reverts to a sharp V at high $Fr$, and the jet becomes slower and thicker. The particle inertia is now too great for widespread early particle ejection; the raft behaves more rigidly, suppressing the wave swell entirely and returning to a V-shaped, capillary-driven collapse.  In conclusion, the enhanced jet seen on intermediate-sized Glaco-treated particles is directly linked to the soft W cavity morphology, and a direct consequence of wettability-driven particle ejection.

\begin{figure*}[t!]
    \begin{subfigure}[t]{1.00\textwidth}   
        \centering
        \vspace{-7mm}
        \input{Figures/Fig7_Jet_Scaling}
    \end{subfigure}
    \begin{subfigure}[t]{0.925\textwidth}   
        \centering
        
        \begin{tikzpicture}[scale=0.925]
            \vspace{1mm}
            \hspace{9.55mm}
            \node[inner sep=0pt] (img) at (0,0) {\includegraphics[width=1\textwidth]{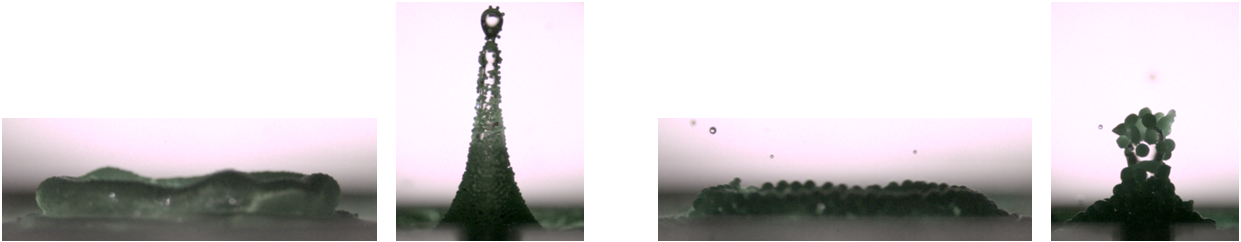}};

            \definecolor{ElasticShade}{RGB}{40,100,180}    
            \definecolor{RigidShade}{RGB}{200,90,40}      
            
            \draw[ElasticShade, line width=2.5pt, opacity = 0.175] ( -8.81, -1.76) rectangle ( -0.475,  1.72);  
            \draw[RigidShade,   line width=2.5pt, opacity = 0.175] (  0.485, -1.76) rectangle (   8.81, 1.72);
            
            \definecolor{Blue}{RGB}{54,80,163}      
            \definecolor{Orange}{RGB}{245,127,34}   
            
            \node[text=Blue, font=\small, anchor=west] at (-8.81+0.06, 1.45) {\text{Elastic regime:}};
            \node[text=Blue, font=\small, anchor=west] at (-8.3, 0.75) {\text{Wave swell}}; 
            \draw[line width=0.2mm, draw=Blue, ->] (-6.80, 0.75) -- (-6.30, 0.75) [anchor=west];
            \node[text=Blue, font=\small, anchor=west] at (-6.25, 0.75) {\text{Tall jet}};
            
            \node[text=Orange, font=\small, anchor=west] at (0.485+0.06, 1.45) {\text{Rigid regime:}}; 
            \node[text=Orange, font=\small, anchor=west] at (1.0, 0.75) {\text{Swell suppressed}}; 
            \draw[line width=0.2mm, draw=Orange, ->] (3.2, 0.75) -- (3.7, 0.75) [anchor=west];
            \node[text=Orange, font=\small, anchor=west] at (3.75, 0.75) {\text{Short jet}};

            \def\xval{-4.30} 
            \def\yval{-0.10}
            \draw[line width=0.5mm, draw=black, -] (\xval, \yval) -- (\xval+0.62, \yval); 
            \node[text=black, font=\footnotesize] at (\xval+0.32, \yval-0.19) {\textbf{2 mm}}; 

            \def\xval{5.02} 
            \def\yval{-0.10}
            \draw[line width=0.5mm, draw=black, -] (\xval, \yval) -- (\xval+0.62, \yval); 
            \node[text=black, font=\footnotesize] at (\xval+0.32, \yval-0.19) {\textbf{2 mm}}; 

            \def\xval{-1.20} 
            \def\yval{1.55}
            \draw[line width=0.5mm, draw=black, -] (\xval, \yval) -- (\xval+0.447, \yval); 
            \node[text=black, font=\footnotesize] at (\xval+0.225, \yval-0.19) {\textbf{2 mm}}; 
            
            \def\xval{8.10} 
            \def\yval{1.55}
            \draw[line width=0.5mm, draw=black, -] (\xval, \yval) -- (\xval+0.447, \yval); 
            \node[text=black, font=\footnotesize] at (\xval+0.225, \yval-0.19) {\textbf{2 mm}}; 

        \end{tikzpicture}
    \end{subfigure}
    \caption{\textbf{ Regime-dependent collapse of Worthington jet height.}
    Collapsed representation of the dimensionless jet height $h^{*}(1+\Pi)$ as a function of Froude number for all particle types and surface treatments. The data separate into two distinct trends corresponding to elastically responsive (upper band, blue) and inertially rigid (lower band, orange) interfacial regimes. Dashed lines indicate power-law fits within each regime, highlighting the systematic difference in scaling behaviour across particle sizes and densities.
    Representative side-view images below illustrate the collapse dynamics associated with these regimes: particle rafts in the elastic regime support wave-swell formation that promotes strong capillary–inertial focusing and tall jets, whereas in the rigid regime swell development is suppressed, resulting in weaker focusing and shorter jets.
    }  
    \label{figure7JetScaling} 
\end{figure*}  
\definecolor{ElasticShade}{RGB}{190,215,245}
\definecolor{RigidShade}{RGB}{245,220,195}

\begin{figure*}[t!]
    \begin{subfigure}[t]{1.00\textwidth}   
        \centering
        \vspace{-2mm}
        \caption{} 
        \begin{tikzpicture}[scale=1.00]
            \node[inner sep=0pt] (img) at (0,0) {\includegraphics[width=1\textwidth]{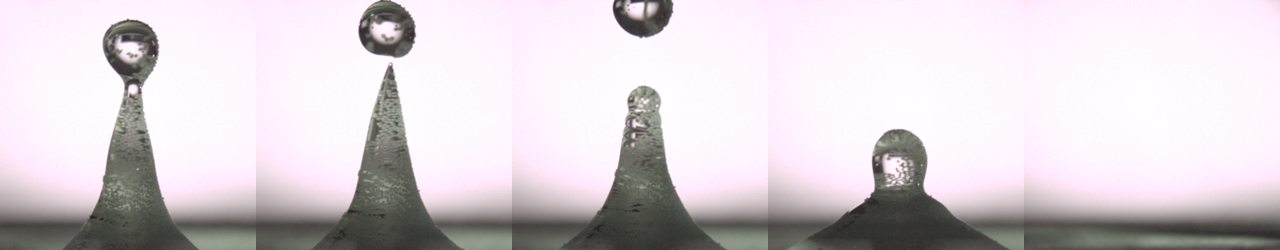}};

            \def\xval{8.00} 
            \def\yval{1.50}
            \draw[line width=0.5mm, draw=black, -] (\xval, \yval) -- (\xval+0.45, \yval); 
            \node[text=black, font=\footnotesize] at (\xval+0.225, \yval-0.19) {\textbf{2 mm}}; 
            
            \node[text=black, font=\small, anchor=center] at (-8.075, 1.5) {\text{$115~\mathrm{\mu m}$ FPE}};

            \node[text=black, font=\small, anchor=center] at (-7.0, -1.95) {\text{$49.47~\mathrm{ms}$}};
            \node[text=black, font=\small, anchor=center] at (-3.5, -1.95) {\text{$52.53~\mathrm{ms}$}};
            \node[text=black, font=\small, anchor=center] at ( 0.0, -1.95) {\text{$55.73~\mathrm{ms}$}};
            \node[text=black, font=\small, anchor=center] at ( 3.5, -1.95) {\text{$64.40~\mathrm{ms}$}};
            \node[text=black, font=\small, anchor=center] at ( 7.0, -1.95) {\text{$103.07~\mathrm{ms}$}};

            \draw[line width=0.4mm, draw=black, ->] (-6.4, 0.7) -- (-6.4, 1.2) [anchor=north];
            \node[text=black, font=\small, anchor=center] at (-6.25, 0.9) {\text{1}};
            
            \draw[line width=0.4mm, draw=black, ->] (-2.9, 1.0) -- (-2.9, 1.5) [anchor=north];
            \node[text=black, font=\small, anchor=center] at (-2.75, 1.2) {\text{1}};
            \draw[line width=0.4mm, draw=black, ->] (-3.1, 0.5) -- (-3.1, 0.0) [anchor=north];

            \draw[line width=0.4mm, draw=black, ->] (0.6, 1.2) -- (0.6, 1.7) [anchor=north];
            \node[text=black, font=\small, anchor=center] at (0.75, 1.4) {\text{1}};
            \draw[line width=0.4mm, draw=black, ->] (0.5, 0.5) -- (0.5, 0.0) [anchor=north];
        \end{tikzpicture}
    \end{subfigure}
    
    \begin{subfigure}[t]{1.00\textwidth}   
        \centering
        \vspace{-2mm}
        \caption{} 
        \begin{tikzpicture}[scale=1.00]
            \node[inner sep=0pt] (img) at (0,0) {\includegraphics[width=1\textwidth]{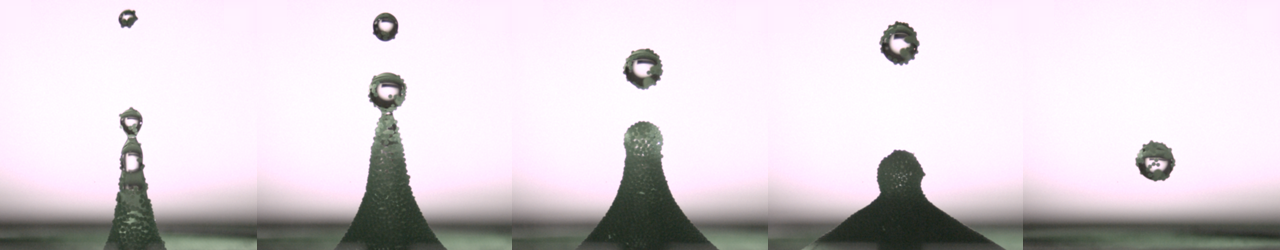}};

            \def\xval{8.00} 
            \def\yval{1.50}
            \draw[line width=0.5mm, draw=black, -] (\xval, \yval) -- (\xval+0.45, \yval); 
            \node[text=black, font=\footnotesize] at (\xval+0.225, \yval-0.19) {\textbf{2 mm}}; 

            \node[text=black, font=\small, anchor=center] at (-8.075, 1.5) {\text{$231~\mathrm{\mu m}$ FPE}};

            \node[text=black, font=\small, anchor=center] at (-7.0, -1.95) {\text{$33.47~\mathrm{ms}$}};
            \node[text=black, font=\small, anchor=center] at (-3.5, -1.95) {\text{$43.60~\mathrm{ms}$}};
            \node[text=black, font=\small, anchor=center] at ( 0.0, -1.95) {\text{$49.60~\mathrm{ms}$}};
            \node[text=black, font=\small, anchor=center] at ( 3.5, -1.95) {\text{$59.73~\mathrm{ms}$}};
            \node[text=black, font=\small, anchor=center] at ( 7.0, -1.95) {\text{$103.87~\mathrm{ms}$}};

            \draw[line width=0.4mm, draw=black, ->] (-6.75, 1.2) -- (-6.75, 1.7) [anchor=north];
            \node[text=black, font=\small, anchor=center] at (-6.6, 1.4) {\text{1}};
            \draw[line width=0.4mm, draw=black, ->] (-6.6, -0.2) -- (-6.6, 0.3) [anchor=north];
            \node[text=black, font=\small, anchor=center] at (-6.45, 0.0) {\text{2}};
            
            \draw[line width=0.4mm, draw=black, ->] (-3.15, 1.1) -- (-3.15, 1.6) [anchor=north];
            \node[text=black, font=\small, anchor=center] at (-3.0, 1.3) {\text{2}};
            \draw[line width=0.4mm, draw=black, ->] (-3.0, 0.2) -- (-3.0, 0.7) [anchor=north];
            \node[text=black, font=\small, anchor=center] at (-2.85, 0.4) {\text{3}};

            \draw[line width=0.4mm, draw=black, ->] (0.5, 0.55) -- (0.5, 1.05) [anchor=north];
            \node[text=black, font=\small, anchor=center] at (0.65, 0.7) {\text{3}};
            \draw[line width=0.4mm, draw=black, ->] (0.5, 0.0) -- (0.5, -0.5) [anchor=north];

            \draw[line width=0.4mm, draw=black, ->] (4.0, 1.35) -- (4.0, 0.85) [anchor=north];
            \node[text=black, font=\small, anchor=center] at (4.15, 1.20) {\text{3}};

            \draw[line width=0.4mm, draw=black, ->] (7.5, -0.25) -- (7.5, -0.75) [anchor=north];
            \node[text=black, font=\small, anchor=center] at (7.65, -0.40) {\text{3}};
        \end{tikzpicture}
    \end{subfigure}
    
    \begin{subfigure}[t]{1.00\textwidth}   
        \centering
        \vspace{-2mm}
        \caption{} 
        \begin{tikzpicture}[scale=1.00]
            \node[inner sep=0pt] (img) at (0,0) {\includegraphics[width=1\textwidth]{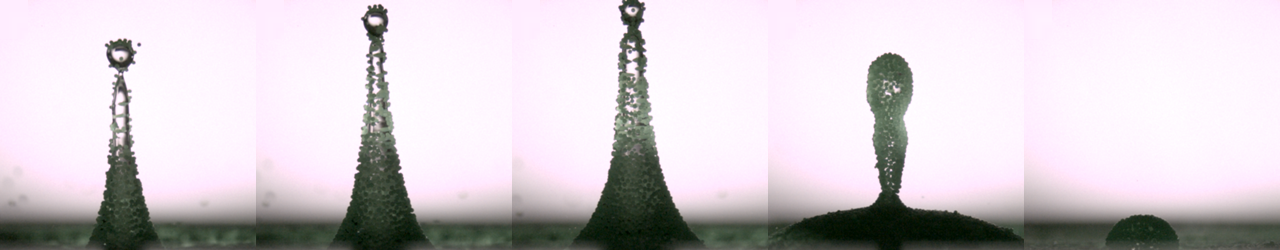}};

            \def\xval{8.00} 
            \def\yval{1.50}
            \draw[line width=0.5mm, draw=black, -] (\xval, \yval) -- (\xval+0.45, \yval); 
            \node[text=black, font=\footnotesize] at (\xval+0.225, \yval-0.19) {\textbf{2 mm}}; 

            \node[text=black, font=\small, anchor=center] at (-7.95, 1.5) {\text{$231~\mathrm{\mu m}$ G-FPE}};

            \node[text=black, font=\small, anchor=center] at (-7.0, -1.95) {\text{$35.60~\mathrm{ms}$}};
            \node[text=black, font=\small, anchor=center] at (-3.5, -1.95) {\text{$41.20~\mathrm{ms}$}};
            \node[text=black, font=\small, anchor=center] at ( 0.0, -1.95) {\text{$48.13~\mathrm{ms}$}};
            \node[text=black, font=\small, anchor=center] at ( 3.5, -1.95) {\text{$75.73~\mathrm{ms}$}};
            \node[text=black, font=\small, anchor=center] at ( 7.0, -1.95) {\text{$498.53~\mathrm{ms}$}};

            \draw[line width=0.4mm, draw=black, ->] (-6.70, 0.7) -- (-6.70, 1.2) [anchor=north];
            \node[text=black, font=\small, anchor=center] at (-6.55, 0.9) {\text{1}};
            
            \draw[line width=0.4mm, draw=black, ->] (-3.25, 1.2) -- (-3.25, 1.7) [anchor=north];
            \node[text=black, font=\small, anchor=center] at (-3.10, 1.4) {\text{2}};

            \draw[line width=0.4mm, draw=black, ->] (0.25, 1.2) -- (0.25, 1.7) [anchor=north];
            \node[text=black, font=\small, anchor=center] at (0.40, 1.4) {\text{3}};
            \draw[line width=0.4mm, draw=black, ->] (0.35, 0.9) -- (0.35, 0.4) [anchor=north];

            \draw[line width=0.4mm, draw=black, ->] (4.0, 0.8) -- (4.0, 0.3) [anchor=north];
            \node[text=black, font=\small, anchor=center] at (5.23, 0.65) {\text{4, Marble-forming jet}};

            \node[text=black, font=\small, anchor=center] at (7.7, -1.00) {\text{4, Liquid Marble}};
        \end{tikzpicture}
    \end{subfigure}

    \begin{subfigure}[t]{1.00\textwidth}   
        \centering
        \vspace{-2mm}
        \caption{} 
        \begin{tikzpicture}[scale=1.00]
            \node[inner sep=0pt] (img) at (0,0) {\includegraphics[width=1\textwidth]{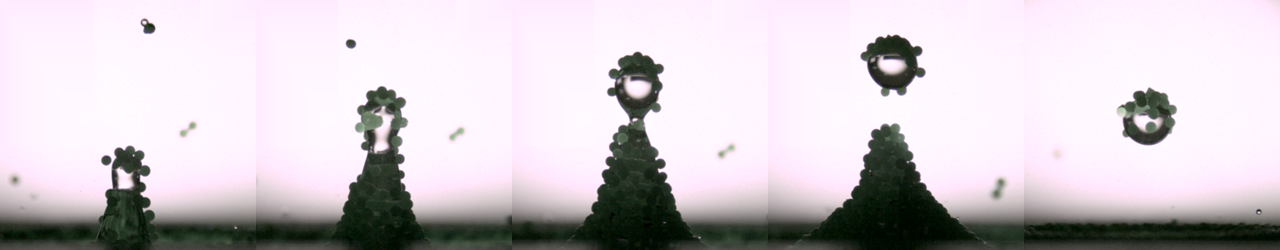}};
            
            \def\xval{8.00} 
            \def\yval{1.50}
            \draw[line width=0.5mm, draw=black, -] (\xval, \yval) -- (\xval+0.45, \yval); 
            \node[text=black, font=\footnotesize] at (\xval+0.225, \yval-0.19) {\textbf{2 mm}}; 

            \node[text=black, font=\small, anchor=center] at (-7.95, 1.5) {\text{$550~\mathrm{\mu m}$ G-FPE}};

            \node[text=black, font=\small, anchor=center] at (-7.0, -1.95) {\text{$33.47~\mathrm{ms}$}};
            \node[text=black, font=\small, anchor=center] at (-3.5, -1.95) {\text{$41.73~\mathrm{ms}$}};
            \node[text=black, font=\small, anchor=center] at ( 0.0, -1.95) {\text{$49.73~\mathrm{ms}$}};
            \node[text=black, font=\small, anchor=center] at ( 3.5, -1.95) {\text{$59.60~\mathrm{ms}$}};
            \node[text=black, font=\small, anchor=center] at ( 7.0, -1.95) {\text{$87.87~\mathrm{ms}$}};

            \draw[line width=0.4mm, draw=black, ->] (-6.50, 1.1) -- (-6.50, 1.6) [anchor=north];
            \node[text=black, font=\small, anchor=center] at (-6.35, 1.3) {\text{1}};
            
            \draw[line width=0.4mm, draw=black, ->] (-3.75, 0.9) -- (-3.75, 1.4) [anchor=north];
            \node[text=black, font=\small, anchor=center] at (-3.60, 1.1) {\text{2}};

            \draw[line width=0.4mm, draw=black, ->] (0.50, 0.3) -- (0.50, 0.8) [anchor=north];
            \node[text=black, font=\small, anchor=center] at (0.65, 0.5) {\text{3}};
            \draw[line width=0.4mm, draw=black, ->] (0.45, 0.0) -- (0.45, -0.5) [anchor=north];

            \draw[line width=0.4mm, draw=black, ->] (4.1, 1.1) -- (4.1, 0.6) [anchor=north];
            \node[text=black, font=\small, anchor=center] at (4.25, 0.95) {\text{3}};

            \draw[line width=0.4mm, draw=black, ->] (7.5, 0.3) -- (7.5, -0.2) [anchor=north];
            \node[text=black, font=\small, anchor=center] at (7.65, 0.15) {\text{3}};
        \end{tikzpicture}
    \end{subfigure}

    \caption{\textbf{Jet-mediated particle ejection and marble formation following cavity collapse. Time-sequence side-view snapshots showing representative modes of particle entrainment by the Worthington jet. Arrows indicate droplet trajectories, and time stamps denote elapsed time after impact.}
    \textbf{(a)} For $115~\mathrm{\mu m}$ FPE untreated particles at $Fr\approx471$, a sharp W-shaped cavity collapse produces a slow, thick cavity jet that pinches off a single large droplet carrying a compact cluster of microplastics. The droplet is rapidly ejected upward and leaves the field of view. 
    \textbf{(b)} For $231~\mathrm{\mu m}$ FPE untreated particles at $Fr\approx314$, a sharp V-shaped collapse generates a fast, thin jet that fragments into three smaller droplets via capillary instability, each transporting only a few particles. Two droplets are ejected at high speed and exit the frame, while the third decelerates and coalesces into the pool. 
    \textbf{(c)} For $231~\mathrm{\mu m}$ G-FPE treated particles at $Fr\approx471$, a soft W-shaped collapse produces an extremely fast and slender jet ejecting three fast tiny droplets that exit the frame. The jet becomes fully armoured by particles, marble-forming jet whose breakup yields a fourth larger, persistent liquid marble which remains intact upon returning to the interface.
    \textbf{(d)} For $550~\mathrm{\mu m}$ G-FPE treated particles at $Fr\approx471$, a sharp V-shaped collapse produces a short, thick jet. Two individual particles are ejected upwards, and a single large droplet pinches off from the jet tip, carrying a small number of particles; this droplet decelerates and subsequently coalesces back into the pool.
    Supplementary Movies 1-4 present the complete time-resolved sequences corresponding to panels (a–d), highlighting the evolution of jet formation and particle entrainment.
    }  
    \label{figure8JetParticle} 
\end{figure*}  

\subsection*{Jet height scaling and regime-dependent collapse}

In this section, we seek a minimal scaling that captures both inertial momentum focusing and the resistance imposed by the particle raft. For a clean interface, the maximum dimensionless jet height follows a power-law dependence on impact inertia, $h^* \propto Fr^\alpha$, reflecting capillary–inertial focusing during cavity collapse\cite{ray2015regimes,michon2017jet,das2022evolution}. The presence of a particle monolayer reduces this focusing efficiency. We account for this additional resistance through a multiplicative attenuation factor $(1+\Pi)^{-1}$, yielding

\begin{equation}
h^{*} \propto \frac{\, \mathrm{Fr}^{\alpha}}{1+\Pi},
\qquad
\Pi = \frac{D_{p}}{D} \, \frac{\rho_{p}}{\rho} \, \left(1+\cos\theta\right),
\end{equation}
where $\Pi$ quantifies the combined geometric, inertial, and wettability-dependent constraint imposed by the raft. Plotting $h^{*}(1+\Pi)$ against $Fr$ on logarithmic axes (Fig.~\ref{figure7JetScaling}) reveals two approximately linear clusters with distinct slopes corresponding to two dynamical regimes of cavity collapse. 
Representative images illustrate the collapse dynamics associated with these regimes. An elastically responsive regime comprises small to intermediate particle rafts ($115-327\,\mathrm{\mu m}$ PE and FPE, treated and untreated), for which the interface remains sufficiently compliant to support wave-swell formation. In contrast, an inertially rigid regime comprises larger and denser rafts ($550-780\,\mathrm{\mu m}$ PE, FPE and glass, treated and untreated), for which the particle layer suppresses swell development and acts as an inertial load during collapse. Hydrophilic glass particles form a special case: their large immersion depths lead to submergence and expulsion from the interface during impact, preventing the formation of a persistent raft. Their dynamics therefore fall outside the present scaling and are excluded. 
Within each regime, the data collapse onto a single power-law relation. In the elastically responsive regime, the jet height scales nearly linearly with inertia, with $\alpha\approx1.03$ ($R^2 \approx 0.85$). In the inertially rigid regime, a steeper dependence is observed, $\alpha\approx1.22$ ($R^2 \approx 0.84$), together with a prefactor lower than 1, indicating diminished vertical momentum focusing.

The separation between regimes reflects the collapse mechanics identified earlier. When the interface remains compliant, energy stored in the wave swell is efficiently redirected into upward jetting, producing near-inertial scaling. When the raft behaves as an inertial load, swell formation is suppressed, and momentum is redistributed into particle acceleration and interfacial dissipation, reducing jet efficiency. Despite the complexity of the interfacial physics, the jet height is therefore captured by a minimal scaling governed by inertial forcing modulated by a single geometric–inertial–capillary resistance.

\subsection*{Jet-mediated particle ejection and marble formation} \label{Sec:JetEjection} 

Beyond impact-driven splashing, a second and often more consequential ejection pathway arises from the collapse of the impact cavity and the subsequent formation of a Worthington jet. Unlike particle splash, which is confined to the earliest stages of impact, jet-mediated ejection can transport particles vertically and generate airborne, particle-laden, droplets of sizes known for its environmental persistence (mist). The efficiency of this pathway is governed primarily by the cavity collapse morphology, which dictates jet velocity, thickness, and breakup behaviour, and by particle–interface characteristics such as the immersion depth, inertia, and wettability. In brief, cavity collapse exhibits a transition from sharp V-shaped to sharp W-shaped morphologies as the Froude number increases for untreated rafts composed of small particles ($115-327\,\mathrm{\mu m}$).  At high $Fr$, sharp W cavities produce a slow and thick jet that pinches off into a single, large tip droplet, often comparable in size to or larger than the original impacting drop. This droplet entrains a dense cluster of microparticles that is ejected vertically (Fig.~\ref{figure8JetParticle}a and Supplementary Movie~1). In this regime, jet-mediated ejection is dominated by the number of particles ($\sim100$) transported per droplet rather than by droplet count. 
At lower $Fr$, the same untreated rafts generate soft W, or sharp V, cavities, leading to fast thin singular jets. These jets undergo capillary breakup into multiple smaller droplets, each typically carrying only a few ($\sim15-30$) particles (Fig.~\ref{figure8JetParticle}b and Supplementary Movie 2). Although individual droplets transport fewer particles, the higher jet droplet count can result in a comparable, or greater, number of ejected particles. For small untreated rafts, at $Fr\approx371$, jet breakup typically ejects $\sim50$ particles per event (Fig.~\ref{figure8JetParticle}b). In contrast, impact-driven splash, i.e. $Fr\approx471$, releases $\sim5-10$ particles (Fig.~\ref{figure3ParticleSplash}a).

Superhydrophobic (Glaco-treated) rafts fundamentally transform the jet-mediated ejection pathway. For small treated particles, i.e. $115-327\,\mathrm{\mu m}$, the cavity collapse takes the shape of a soft W, or a flat V, across the explored $Fr$ range. This produces extremely fast and thin jets which, at early times, eject high-speed microdroplets that can transport significant particle loads into the air. As the jet develops, particles rapidly migrate and accumulate along the liquid–air interface, ultimately forming a continuous particle shell around the liquid column. This process gives rise to a fully armoured, marble-forming jet (Fig.~\ref{figure8JetParticle}c and Supplementary Movie 3). There are two critical consequences of these dynamics: first, capillary breakup of the armoured jet produces discrete droplets that remain fully coated, yielding stable liquid marbles that resist coalescence upon re-impact with the pool surface \cite{aussillous2001liquid,aussillous2006properties}. Second, the particle layer fundamentally alters the interfacial dynamics of the jet, enabling it to sustain high velocities while undergoing vigorous fragmentation. Together, these effects make the marble-forming jet a highly efficient ejection mechanism. It combines a high particle load per droplet with the production of numerous carrier droplets,  thereby generating a dense population of persistent airborne liquid marbles. Untreated hydrophobic PE and FPE rafts can also generate particle-laden droplets via the jet breakup; however, particle coverage in these cases remains partial rather than forming a continuous shell. These partially armoured droplets lack the non-wetting shell characteristic of true liquid marbles and frequently coalesce upon falling back to the pool, limiting their atmospheric lifetime compared with the fully armoured droplets produced by treated rafts.

For larger particles, i.e. $550-780\,\mathrm{\mu m}$, jet-mediated ejection is strongly suppressed regardless of wettability. In this regime, cavity collapse is consistently forced towards a sharp V-shaped geometry across the explored $Fr$ range, as the inertia of the raft suppresses wave swell formation and vertical momentum focusing. Although the interface remains covered by particles, their large inertia inhibits jet acceleration and thinning. The resulting Worthington jet is short, thick, and flat, exhibiting weak vertical extension and limited breakup. Rather than fragmenting into multiple droplets, the jet produces at most a single, weak pinch-off event: a small number of individual particles may be accelerated upward, and a single particle-laden droplet can detach from the jet tip (Fig.~\ref{figure8JetParticle}d and Supplementary Movie 4). Consequently, jet-mediated ejection becomes inefficient as particle inertia dominates. The behaviour of glass spheres provides a stringent confirmation of this hierarchy of governing parameters. Despite superhydrophobic treatment, their high density ($\rho_P\approx2.50\,\mathrm{g\,cm^{-3}}$) prevents significant particle ejection during the jet phase. Their inertia imposes a severe load on the jet, suppressing jet height, thinning, and droplet formation, and thereby inhibiting aerosolisation through both impact- and jet-driven pathways.

Together, these results establish jet-mediated ejection as the dominant pathway for the atmospheric release of microplastics from particle rafts, particularly for small, weakly immersed, superhydrophobic particles. By directly linking cavity collapse morphology, jet dynamics, and particle armouring, this framework explains why such particles present the highest risk for long-range atmospheric transport via the formation of long-lived liquid marble aerosols.

\section*{Discussion}

This study has established a mechanistic framework linking droplet impact dynamics to microplastic aerosolisation from floating particle rafts. We show that particle monolayers fundamentally control both splashing and cavity collapse. In brief, the splash onset is governed not by any single material parameter but by a composite geometric–inertial–capillary resistance, captured by $\Lambda = (D_p/D)(\rho_p/\rho)(1-\cos\theta)^3$, which collapses the spreading–splash transition across all materials and treatments. Small ($115$–$327~\mathrm{\mu m}$), deeply immersed particles ($h\gtrsim0.41$) stabilise the spreading lamella, shifting the splash onset to higher Weber numbers and restricting ejection to microdroplets. Conversely, larger ($550$–$780~\mathrm{\mu m}$), or weakly immersed, particles ($h\lesssim0.11$) protrude above the interface, introduce effective geometric roughness, destabilise the rim, and induce fingering that culminates in violent splashing at reduced Weber numbers.

Beyond the splash onset, we have identified jet-mediated ejection as the dominant pathway for particle transfer. Cavity collapse morphology transitions from wave-swell-mediated W-shaped closure to V-shaped collapse as particle size and inertia increase, suppressing capillary–inertial focusing. 
Despite this complexity, the maximum jet height is well described by $h^{*} \propto \frac{\mathrm{Fr}^{\alpha}}{1+\Pi}$, where $\Pi = (D_p/D)(\rho_p/\rho)(1+\cos\theta)$ represents an effective geometric–inertial resistance imposed by the raft. When rescaled by $(1+\Pi)$, the data separate into two distinct regimes: an elastically responsive regime ($115-327~\mathrm{\mu m}$) exhibiting near-inertial scaling ($\alpha\approx1$), and an inertially rigid regime ($550-780~\mathrm{\mu m}$) characterised by steeper scaling and reduced momentum focusing.

Treated (superhydrophobic) rafts enable a qualitatively distinct ejection mode. For small particles ($115$–$327~\mathrm{\mu m}$), weak capillary adhesion permits immediate particle detachment upon impact, while subsequent cavity collapse produces particle-armoured “marble-forming” jets that fragment into persistent liquid marbles resistant to coalescence. In contrast, untreated rafts exhibit limited particle ejection during impact; particles are instead primarily entrained within the rising jet, forming only partially armoured droplets that frequently coalesce upon re-impact. Fully armoured droplets from superhydrophobic rafts combine high particle load with extended atmospheric lifetime, representing one of the most efficient aerosolisation pathways observed.

Together, these findings reveal how particulate monolayers transform canonical droplet-impact physics, introducing interfacial resistance that governs both splash thresholds and vertical momentum concentration. By linking collapse morphology, jet scaling, and particle entrainment within a unified framework, this work provides a predictive basis for quantifying ocean-to-atmosphere microplastic transfer under rainfall and extends classical impact theory to particle-laden and contaminated interfaces.

\section*{Methods} 

\subsection*{Experimental Setup}
The experimental apparatus was designed for the optimum visualisation of the interplay between a falling droplet under gravity and a floating particle raft on a water pool (Fig.~\ref{figure1setup}a). Droplets were generated using an AL-300 Aladdin InfusionONE syringe pump operated at a constant flow rate of 20 $\mathrm{\mu L\,min^{-1}}$. A blunt-end stainless-steel dispensing needle (Gauge 20, inner diameter, $0.61\,\mathrm{mm}$) was used to form pendant droplets of deionised water, which were detached under their own weight to produce droplets of diameter $D =2.55\,\mathrm{mm}$. The impact velocity, $U$, was systematically varied between $0.39$ and $3.46\,\mathrm{m\,s^{-1}}$ by adjusting the height of the dispensing needle using a precision mechanical vertical stage.

A rectangular acrylic tank (internal dimensions: $65\times49\times85\,\mathrm{mm}$) was used as the impact pool. A concave meniscus was generated by leaving a small clearance below the rim when filling the tank with deionised water. This configuration produced a stable inward-sloping free surface that promoted migration and confinement of the particle raft towards the centre. A depth to droplet diameter ratio of $H/D \approx33$ ensures negligible bottom effects, and the pool can therefore be considered semi-infinite. The lateral confinement ratio ($R/D\approx10-13$) is likewise sufficiently large to minimise early-time wall reflections, as surface wave propagation from the impact site to the walls occurs on a timescale much longer than that of cavity collapse and Worthington jet formation.

Two synchronised high-speed cameras were used to resolve the droplet impact dynamics from vertically offset side views centred on the impact plane (Fig.~\ref{figure1setup}b).
A Phantom V710 high-speed camera (1280 $\times$ 800 pixels) equipped with a Navitar $12\times$ lens recorded the upper region at 7,500 fps with an exposure time of $60\,\mathrm{\mu s}$, resolving the splashing, particle ejection, and the formation of the subsequent Worthington jet. Simultaneously, a Phantom TMX high-speed camera (1280 $\times$ 800 pixels) fitted with the same lens recorded the lower region at 7,500 fps with an exposure time of $20\,\mathrm{\mu s}$, resolving crater formation within the particle raft and subsequent cavity collapse. Both views were illuminated using a Photofluor II lamp (89 North), providing diffuse front lighting. The spatial resolutions were $55$ and $86\,\mathrm{pixels\,mm^{-1}}$, for the upper and lower views, respectively. All experiments were performed under standard ambient laboratory conditions at a temperature of $22\pm1\,^\circ C$ and a relative humidity of $45 - 55\%$.

\subsection*{Particle Positioning} 
The microplastic particles were gently deposited onto the water surface by lightly tapping the storage container to release a controlled amount of particles. A rubber suction bulb was then used to blow gentle air streams across the surface, enabling fine adjustment of particle positions and preventing overlap, particularly for the smaller particles that tended to cluster. To achieve a closely packed monolayer, the tank walls were lightly tapped to generate surface waves that promoted self-assembly into a uniform two-dimensional hexagonal packing. This procedure was repeated until a continuous and tightly packed particle raft was formed. Following each droplet impact, additional particles were introduced to restore full surface coverage, and the arrangement was fine-tuned using the same procedure. Real-time feedback from the high-speed cameras was used to verify the uniformity and ensure that each droplet impacted the centre of a fully packed raft, maintaining consistency and reproducibility across all experiments.




\subsection*{Materials} 

Deionised water was obtained from a Milli-Q purification system (Millipore, resistivity $18.2\,\mathrm{M\Omega\cdot cm}$). Its physical properties at $22\,\pm$\,1\,$\mathrm{^\circ C}$ were a density of $998\,\mathrm{kg\,m^{-3}}$, dynamic viscosity of $1.0\,\mathrm{mPa\,s}$, and surface tension of $72.8\,\mathrm{mN\,m^{-1}}$. 

The experiments utilised spherical microparticles of various materials, sizes, and densities to form the floating rafts. Polyethylene microspheres (Cospheric LLC, USA) were used in two density grades: fluorescent green polyethylene (FPE) ($\rho_P=1.02\,\mathrm{g\,cm^{-3}}$) with mean diameters of 115, 231, 327, 550, 780, and 925\,$\mathrm{\mu m}$; and white polyethylene (PE) ($\rho_P=1.35\,\mathrm{g\,cm^{-3}}$) with mean diameters of 327 and $780\,\mathrm{\mu m}$. Hydrophilic glass microspheres (Glass) ($\rho_P=2.50\,\mathrm{g\,cm^{-3}}$, $D_P=625\,\mathrm{\mu m}$) were sourced from Thermo Fisher Scientific. For all particle types, the reported diameters correspond to the mean values of the supplier’s nominal size distributions. All particles were used as received, without any initial surface modification. Selected batches of each particle type were rendered superhydrophobic by coating their surfaces with Ultra Ever Dry Glaco (Soft99 Co., Japan) to investigate the influence of surface wettability. The Glaco-treated (G-FPE, G-PE, and G-Glass) particles were left to dry under ambient laboratory conditions for 24 hours, producing a thin layer of hydrophobic nanoparticles that imparted a superhydrophobic finish.

\subsection*{Profilometer}
The wettability of each microparticle type was quantified from its equilibrium immersion depth at the air–water interface. Multiple spheres of the same particle type were gently deposited, and their height above the quiescent water surface was measured using a vertical-scanning optical profilometer (InfiniteFocus, Alicona GmbH) (Fig.~\ref{figure1setup}c). The submerged depth was then defined as $h_{sub} = D_p-h_{above}$, where $D_p$ is the particle diameter and $h_{above}$ is the measured height of the sphere above the undisturbed interface (Fig.~\ref{figure1setup}d). The dimensionless immersion ratio was defined as $h=h_{sub}/D_p$. Assuming a negligible Bond number (consistent with all particles used), the apparent contact angle was obtained from the geometric relation for a sphere intersecting a locally flat interface, $\theta=cos^{-1}(1-2h)$. The reported values represent the mean $\pm$ standard deviation across all spheres measured for each particle type.




\bibliography{Bibliography}

\section*{Acknowledgments} 
This work was supported by a UCL STEPS Grant (Project 586921). 
M.A.Q-S. acknowledges support from DGAPA through Subprograma de Incorporación de Jóvenes Académicos de Carrera (SIJA) and the grant PAPIIT-UNAM IA101025; and support from the CIC through the Programa de Movilidad Académica Nacional e Internacional para el Subsistema de la Investigación Científica.

\section*{Author contributions}
M.H.I., A.A.C.P., J.R.C.P., and  M.A.Q-S. conceived the project. 
M.H.I., and  M.A.Q-S designed the experiments.
M.H.I., and  M.A.Q-S carried out the experiments. 
M.H.I., carried out the analysis. 
M.H.I., and  M.A.Q-S. led the writing.
A.A.C.P., and J.R.C.P. reviewed the manuscript. 
J.R.C.P. and  M.A.Q-S. lead the project. 

\section*{Data availability}
The experimental raw data supporting the findings of this study are
available from the corresponding author upon request. The processed
data are available within the manuscript and its Supplementary Information Files. Source data are provided along with this paper.
The datasets generated and/or analysed during the current study are available in the Zenodo repository, [Link to Dataset]. The files will be created after the manuscript is accepted.

\section*{Competing interests}
The authors declare no competing interests.

\section*{Publisher’s note}
Springer Nature remains neutral with regard to jurisdictional claims in published maps and institutional affiliations.

\section*{Additional information}

\textbf{Supplementary Information} The online version contains supplementary material available at [DOI link to be created by Nature Publishing Group].


\end{document}